\documentclass[11pt,journal,onecolumn]{IEEEtran}

\usepackage{cite}
\usepackage{graphicx}
\graphicspath{{./fig_R1/}}

\usepackage{amsmath}
\usepackage{amssymb}
\usepackage{amsfonts}
\usepackage{bbm}
\usepackage{bm}
\usepackage{lipsum}
\usepackage{array}
\usepackage{placeins}
\usepackage{mdwmath}
\usepackage{mdwtab}
\usepackage{eqparbox}
\usepackage[font=footnotesize]{subfig}
\usepackage{fixltx2e}
\usepackage{stfloats}
\usepackage{multirow}
\usepackage[]{algorithm}
\usepackage[noend]{algpseudocode}

\makeatletter
\def\BState{\State\hskip-\ALG@thistlm}
\makeatother

\fnbelowfloat

\ifCLASSOPTIONcaptionsoff
  \usepackage[nomarkers]{endfloat}
 \let\MYoriglatexcaption\caption
 \renewcommand{\caption}[2][\relax]{\MYoriglatexcaption[#2]{#2}}
\fi
 \let\MYorigsubfloat\subfloat
 \renewcommand{\subfloat}[2][\relax]{\MYorigsubfloat[]{#2}}

\usepackage[hidelinks]{hyperref}
\usepackage{url}

\usepackage[]{algorithm}
\usepackage[noend]{algpseudocode}

\renewcommand{\theequation}{\arabic{equation}}
\newcommand{\be}{\begin{equation}}
\newcommand{\ee}{\end{equation}}
\newcommand{\bea}{$$ \begin{array}{lll}}
\newcommand{\eea}{\end{array} $$}
\newcommand{\bi}{\begin{itemize}}
\newcommand{\ei}{\end{itemize}}

\newtheorem{definition}{Definition}[section]

\def\l0{\ell^0(\mathbb{Z}^{2})}

\def\RR{\rm \hbox{I\kern-.2em\hbox{R}}}
\def\NN{\rm \hbox{I\kern-.2em\hbox{N}}}
\def\ZZ{\rm {{\rm Z}\kern-.28em{\rm Z}}}

\def\<{\langle}
\def\>{\rangle}

\def\Chi{\raise .3ex \hbox{\large $\chi$}}

\def\v1{\tilde v}

\def\[{\Bigl [}
\def\]{\Bigr ]}
\def\({\Bigl (}
\def\){\Bigr )}
\def\[{\Bigl [}
\def\]{\Bigr ]}
\def\({\Bigl (}
\def\){\Bigr )}

\def\l{\lambda}

\ifCLASSINFOpdf
\else
\fi

\usepackage{epstopdf}
\begin{document}
\title{A Novel Ridge Detector for Nonstationary Multicomponent Signals: Development and Application to Robust Mode Retrieval}

\author{Nils Laurent, Sylvain Meignen}

\markboth{IEEE Transactions on Signal Processing }%
{Shell \MakeLowercase{\textit{et al.}}: Bare Demo of IEEEtran.cls for Journals}

\maketitle

\begin{abstract}
Time-frequency analysis is often used to study non stationary multicomponent signals, which can be viewed as the surperposition of modes. 
To understand such signals, it is essential to identify the ridges associated with the modes in the time-frequency plane. 
As existing ridge detectors are often not enough robust to noise, we here develop a novel approach to ridge detection 
based on the gathering of ridge portions in the time-frequency plane, which we coin RRP-RD. 
Such a technique is proved to be much more robust to noise than state-of-the-art methods based on the same framework, and we also demonstrate  
its benefits for mode retrieval. 
\end{abstract}

\begin{IEEEkeywords}
AM/FM multicomponent signals, Short-time Fourier transform, Ridge detection, 
Mode retrieval.
\end{IEEEkeywords}

\IEEEpeerreviewmaketitle
\section{Introduction}
\label{sec:introduction}
\IEEEPARstart{M}{any} nonstationary signals such as audio signals (music, speech, bird songs) \cite{gribonval2003harmonic}, 
electrocardiogram \cite{Herry2017}, thoracic and abdominal movement signals \cite{Lin2016}, 
can be modeled as a superposition of amplitude- and frequency-modulated (AM/FM) modes. Such signals are called \emph{multicomponent signals} (MCSs), 
and  \emph{time-frequency} (TF) analysis is often used to deal with them \cite{Flandrin1998,Boashash2003,stankovic2014time}, essentially 
because the modes are associated with curves in the TF plane, called \emph{ridges}. 
Several techniques were developed for ridge detection using the idea that the ridges 
correspond to \emph{local modulus maxima along the frequency axis} (LMMFs) of some \emph{time-frequency representation} (TFR). 
It was shown in \cite{stankovic2001measure,stankovic2001performance} that, when the TFR is the spectrogram, the locations of the LMMFs in 
the TF plane are estimates of the \emph{instantaneous frequencies} (IF) of the modes, the quality of estimation depending on the noise level and on the
length of the analysis window. Still for the purpose of IFs estimation, ridge detection has been applied to many other TFRs than the spectrogram, 
such as the \emph{continuous wavelet transform} (CWT) \cite{Carmona1997}, the \emph{short-time Fourier transform} (STFT)  
\cite{meignen2018retrieval}, or the  \emph{Wigner-Ville distribution} (WVD)  \cite{djurovic2004algorithm}.

The studies on the quality of IFs estimation using TF ridges often assume a low noise level \cite{stankovic2001measure}, but 
at high noise level, the LMMFs that define the ridges in the noiseless case may no longer exist. 
To define IF estimates in such cases, an algorithmic approach developed in \cite{djurovic2004algorithm}, and based on WVD, exploited  the ideas that 
the IFs of the modes vary smoothly and correspond, in the TF plane, to coefficients with large WVD magnitude. 
However, this last assumption is only valid to a certain extent since a high noise can generate zeros of the TFR at expected IFs TF location.
We shall here also mention that there exist alternative techniques to extract the ridges in the TF plane not specifically based on LMMFs but 
using optimization procedures instead \cite{Carmona1999,zhu2019two}.  These \emph{ridge detectors} (RDs) depend on an initial guess for the ridges, called \emph{skeleton of the transform}, which is however very hard to obtain in heavy noise situations.  

Another important application of RDs is mode retrieval for which many different techniques were developed based on different types of TFR. 
A mode retrieval technique from CWT  was proposed in  \cite{Carmona1997}, while from synchrosqueezed CWT (SST) in \cite{Daubechies2011}. 
When STFT is used as TFR,  one may refer to \cite{meignen2018retrieval},  and also to \cite{laurent2020novel} in which 
the authors use a local linear chirp approximation to improve mode retrieval.  As IFs estimation using STFT ridges may be drastically altered when the modes 
are interfering in the TF plane, thus hampering mode retrieval, several techniques based on \emph{adaptive short-time Fourier transform} were  recently developed to reconstruct the 
modes \cite{li2019adaptive}. A discrete  version of this variant of STFT, known as the \emph{signal separation operator}, was introduced and then used for mode reconstruction in \cite{chui2016signal}, and then further developed in \cite{li2020direct}, assuming a linear chirp approximation for the modes.  Note that the mathematical analysis of the latter technique is available in 
\cite{chui2021analysis}. Finally, one shall also mention mode retrieval techniques based  on synchrosqueezed STFT (FSST) as proposed in 
\cite{Thakur2011,Thakur2013,oberlin2014}. Though very interesting, the robustness of all these mode retrieval techniques remains to be investigated in 
very noisy situations. 

Our goal in this paper is to propose a new RD based on STFT that is competitive in very noisy situations.  
For that purpose, we first explain how LMMFs can be linked by considering the notion of 
\emph{relevant ridge portions} (RRPs) we introduce. These are then gathered together exploiting some specific structures called \emph{basins of attraction} associated with RRPs \cite{Meignen2016a}, and the ridges are finally defined  from RRPs using a spline least-square approximation.
In the definition of our new RD we make the assumption that the modes are not crossing, though to deal with such situations 
seems feasible by imposing regularity constraints on the extracted modes \cite{Carmona1999}, or by analyzing the signal in the time domain using a parametric approach \cite{chen2017intrinsic}. 

The paper is organized as follows: in Section \ref{sec:notations}, we first introduce basic notations on  MCSs, STFT, and on the most commonly 
used STFT-based RD. Then, we explain how the latter can be made more adaptive by using 
a local chirp rate estimate, as recently proposed in \cite{colominas2020fully}. The description of the proposed new RD is  
carried out in Section \ref{sec:defRD}, and is followed by two sections, the first one describing a technique to estimate the number of modes and the other exploiting notions introduced 
in Section \ref{sec:defRD} for the purpose of mode retrieval. 
Section \ref{sec:res} is then devoted to the comparison of the proposed new RD with state-of-the-art TF-based RDs, highlighting the improvement brought 
by the former in heavy noise situations, and mode reconstruction is then discussed. An application to the analysis of gravitational-wave concludes the paper.

\section{Definitions and Notations}
\label{sec:notations}
\subsection{Multicomponent Signal Definition}
\label{sec:def_MCS}
In this paper, we will study MCSs defined as a superposition of AM/FM components (or modes):
 \begin{equation}
 \label{def:MCS}
 f[n] = \sum \limits_{p=1}^P f_p[n]~~~~\textrm{with}~~
 f_p[n] = A_p[n] e^{i 2\pi \phi_p[n]},
\end{equation}
for some finite $P \in \mathbb{N}$, $A_p[n]$ and $\phi'_p[n]$ 
being respectively the \emph{instantaneous amplitude} (IA) and IF of 
$f_p$ satisfying: $A_p[n]>0,\phi'_p[n]>0$ and 
$\phi'_{p+1}[n]>\phi'_p[n]$ for each time index $n$.
We also assume that $A_p$ is differentiable with $|A_p'[n]|$ small, that the modes are separated with resolution 
$\Delta$ and their modulations are bounded by $B_f$. The last two conditions mean that for each time index $n$,
\begin{equation}
\begin{aligned}
\label{def:sep}
\forall \ 1 \leq p \leq P-1, \  \phi_{p+1}'[n] - \phi_{p}'[n] > 2 \Delta\\
\forall \ 1 \leq p \leq P,  \ |\phi_p''[n]| \leq B_f.
\end{aligned}
\end{equation}
\subsection{Short-Time Fourier Transform}
\label{sec:STFT}
Let $\tilde{f}$ be a complex discrete signal of length $L$ altered
by a complex additive noise $\varepsilon$, and such that $\tilde{f}[n] = \tilde{f}(\frac{n}{L})$:
\begin{equation}
\label{def:ftilde}
 \tilde{f} := f + \varepsilon,
\end{equation}
and $g$ a discrete real window supported on $[-\frac{M}{L},\frac{M}{L}]$. The STFT of $\tilde f$ is defined as follows:

\begin{equation}
\label{def:STFT}
V_{\tilde{f}}^{g} [m,k] := \sum_{n = 0}^{N-1} \tilde{f}[n + m - M]g[n - M] e^{-2i\pi \frac{k}{N}(n - M)},
\end{equation}
with $2M+1\leq N$, where $N$ is the number of frequency bins, and the index $k$ 
corresponds to the frequency $k \frac{L}{N}$, $-\frac{N}{2} \leq k \leq \frac{N}{2}-1$, to fullfil Nyquist frequency constraint. 
 The STFT is invertible, provided $g[0] \neq 0$, since one has:
\begin{equation}
\label{def:inv_STFT}
\tilde{f}[n] = \frac{1}{g[0]N} \sum_{k=-\frac{N}{2}}^{\frac{N}{2}-1} V_{\tilde{f}}^{g}[n,k].
\end{equation}
Note that, as we are going to deal with MCSs of type  \eqref{def:MCS},  $V_{\tilde{f}}^{g}[n,k]$ is with very low amplitude when $k < 0$. 
When $\tilde f$ is real valued (meaning both $f$ and $\varepsilon$ are real), the reconstruction formula reads:
\begin{equation}
\label{def:inv_STFT1}
\tilde{f}[n] = \frac{2}{g[0]N} \Re \{  \sum_{k=0}^{\frac{N}{2}-1} V_{\tilde{f}}^{g}[n,k] \},
\end{equation}
with $\Re \{ X \}$ the real part of complex number $X$. 
\subsection{Classical STFT-based RD}
\label{sec:ridge}
 The most commonly used RD was introduced by Carmona et \emph{al.}  \cite{Carmona1997} to compute the ridges of CWT 
 and can easily be adapted to STFT. It consists of finding the $P$ ridges associated with the modes in the TF plane by computing: 
\begin{equation}
\label{eq:Optimization_discrete}
\begin{aligned}
\max_{\bm{\varphi}} \sum_{\tiny \begin{array}{c}1 \leq p \leq P\\  0\leq n  \leq L-1 \end{array}}
| V_{\tilde f}^g \left [ n,\varphi_p[n] \right ] |^2 \\
- \alpha ( \frac{\Delta^1 \varphi_p[n]L^2}{N})^2 - & \delta ( \frac{\Delta^2 \varphi_p[n]L^3}{N} )^2,
\end{aligned}
\end{equation}
with $\bm{\varphi} = (\varphi_p)_{p=1,\cdots,P}$ where $\varphi_p : \{0,\cdots,L-1\} \mapsto \{0,\cdots,\frac{N}{2}-1\}$, $\alpha$ and $\beta$ are both positive, and in which 
$\frac{\Delta^1 \varphi_p[n] L^2}{N} = \frac{(\varphi_p[n+1] - \varphi_p[n])L^2}{N}$
and
$\frac{\Delta^2 \varphi_p[n]L^3}{N} = \frac{(\varphi_p[n+1] - 2\varphi_p[n] + \varphi_p[n-1])L^3}{N}$ are estimates of $\phi''_p[n]$ and $\phi'''_p[n]$.
To consider penalization terms is however not relevant when the IFs of the modes actually correspond to LMMFs, which is the 
case at low noise level and, as the choice for $\alpha$ and $\delta$ drastically alters ridge detection in a noisy context, 
penalization terms are often not considered in \eqref{eq:Optimization_discrete} \cite{meignen2017demodulation}. 
Alternatively, one can use the bound $B_f$ on the frequency modulation to extract a first ridge, and then replace \eqref{eq:Optimization_discrete} 
by a \emph{peeling algorithm}. In a nutshell,  a first ridge is extracted as follows \cite{colominas2020fully}:
\begin{equation}
\label{eq:Optimization_bound}
\max_{\varphi_1} \sum_{n = 0}^{L-1} | V_{\tilde f}^g \left [n,\varphi_1[n] \right ] |^2, \quad \text{s.t. } |\Delta^1 \varphi_1[n]| \frac{L^2}{N} \leq B_f.
\end{equation}
Then, after $\varphi_1$ is computed, one defines $V_{\tilde f,0}^g := V_{\tilde f}^g$, and RD continues replacing $V_{\tilde f,0}^g$ by:
\begin{align}
\label{def_x1}
V_{\tilde f,1}^g [n,k] :=
\left \{\begin{array}{l}
0 \text{, if } |k - \varphi_1[n]| \leq   \frac{\Delta N}{L} \\
V_{\tilde f,0}^g [n,k] \text{, otherwise.}
\end{array}
\right .
\end{align}
This enables the computation of $\varphi_2$ replacing $V_{\tilde f}^g$ 
by $V_{\tilde f,1}^g$ in \eqref{eq:Optimization_bound}, and then the definition of
$V_{\tilde f,2}^g$ replacing $V_{\tilde f,0}^g$ by $V_{\tilde f,1}^g$ and $ \varphi_1$ by $\varphi_2$ in \eqref{def_x1}. 
Such a procedure is iterated until $P$ ridges $(\varphi_p)_{p=1,\cdots,P}$ are extracted.

In practice, to implement \eqref{eq:Optimization_bound}, one first considers an initial time index $n_0$, then defines    
\begin{equation}
    \label{eq:max_magnitude_freq}
    k_0 := \mathop{\textrm{argmax}}_{0 \leq k \leq N-1} |V_{\tilde f}^g [n_0,k]|,
\end{equation}
and finally sets $\varphi_1[n_0] := k_0$. To define $\varphi_1$ on 
$\{n_0+1,\cdots,L-1\}$, one then uses the following recurring principle 
starting from $n = n_0$:
\begin{equation}
    \label{eq:mb_interval0}
    \varphi_1[n+1] := \mathop{\text{argmax}}_{k} \left \{ |V_{\tilde f}^g[n+1,k]|,  
    | k -\varphi_1[n]| \leq  \frac{NB_f}{L^2}  
    \right \}.
\end{equation}
The same principle is applied on $\{0,\cdots,n_0-1\}$, starting from  $n = n_0$ and replacing $n+1$ by $n-1$ in \eqref{eq:mb_interval0}.
Finally, the procedure is run again starting from other initial time indices  
to define other candidates for $\varphi_1$, and the ridge finally kept 
among all the candidates is the one maximizing the energy in the TF plane, 
i.e. $\sum\limits_n |V_{\tilde f}^g [n,\varphi_1[n]]|^2$.
This RD will be called \emph{Simple Ridge Detection} (S-RD) in the sequel. 

\begin{figure}[!htb]
\centering
\begin{minipage}{0.5\linewidth}
	\begin{tabular}{c}
	\includegraphics[width=6cm, height=6cm] {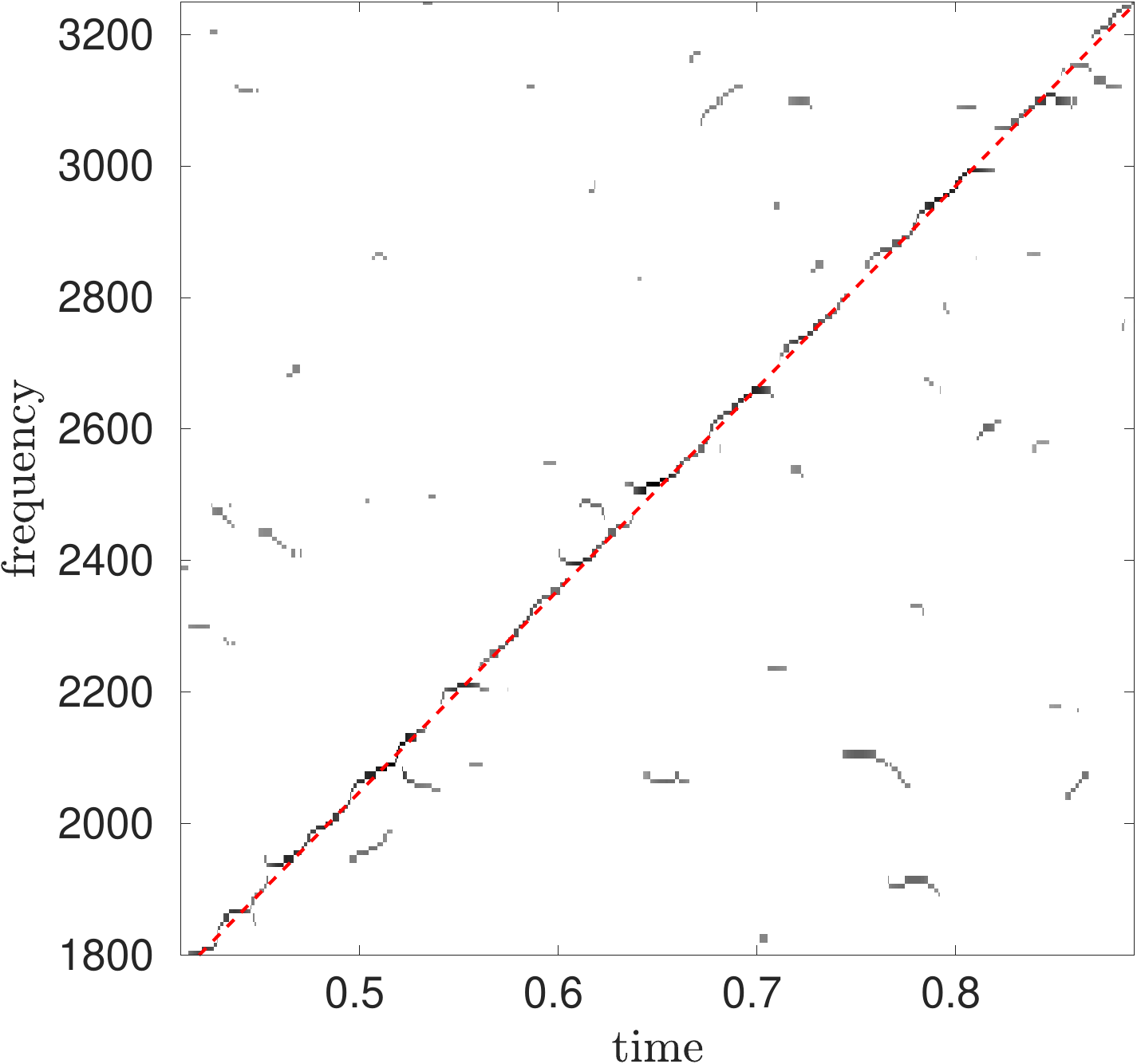}\\
	 \hspace{0.6 cm} 
	\end{tabular}
\end{minipage}%

\caption{LMMFs corresponding to the three largest STFT modulus maxima for each time instant along with the true IF (dashed line) of the linear chirp (SNR= -10 dB )}
\label{Fig1}
\end{figure}

There are however two strong limitations to S-RD. 
The first one is that  each ridge is built by chaining LMMFs in the TF plane assuming 
the chain is continuous. However in heavy noise situations, zeros of STFT may appear at TF locations corresponding the IFs of the modes, 
resulting in the splitting of a chain of LMMFs into two chains of LMMFs at these locations.  
This is illustrated on a linear chirp in Fig. \ref{Fig1} in which we display the magnitude of the 
LMMFs associated with the three largest STFT modulus maxima at each time instant, along with the true IF (SNR = -10 dB). 

The second important drawback of S-RD is that the jumps allowed between two successive time indices depend on the  modulation parameter 
$B_f$  which is fixed a priori and positive, and thus this method does not adapt to the local variations of the frequency modulation of the modes. 
In this regard, we recall, in the following section, how to introduce some kind of adaptivity in 
RD, as proposed in \cite{colominas2020fully}, by removing the dependency of S-RD on the modulation parameter $B_f$.

\subsection{Adaptive Ridge Detection}
\label{sec:adpative_ridge}
To circumvent the lack of adaptivity of S-RD to the local variations of the frequency modulation of the modes, 
a novel approach called \emph{modulation based ridge detection} (MB-RD) was proposed 
in \cite{colominas2020fully}. In a nutshell, this approach considers the following complex  modulation operator used in the definition of the 
\emph{second order synchrosqueezing transform} \cite{behera2018theoretical}:
\begin{equation}
\label{def:secondorder}
\tilde q_{\tilde f}[n,k]
=
\frac{1}{2i \pi} \frac{V_{\tilde f}^{g''}[n,k] V_{\tilde f}^g[n,k]- (V_{\tilde f}^{g'}[n,k])^2}
{V_{\tilde f}^{tg}[n,k] V_{\tilde f}^{g'}[n,k] - V_{\tilde f} ^{tg'}[n, k] V_{\tilde f}^g[n, k]},
\end{equation}
in which $V_{\tilde f}^{g'}, V_{\tilde f}^{tg}, V_{\tilde f}^{g''}, V_{\tilde f}^{tg'}$ 
are respectively the STFTs of $\tilde f$ computed with windows $n \mapsto g'[n], (tg)[n], g''[n]$ and $(tg')[n]$. 
It is shown in \cite{behera2018theoretical} that $\hat q_{\tilde f}[n,k] = \Re \left \{ \tilde q_{\tilde f}[n,k] \right \}$, consists of an estimate of the frequency modulation of the closest mode to 
$[n,k]$ in the TF plane.  To extract the first ridge, MB-RD uses the same recurring principle 
as S-RD introduced in Section \ref{sec:ridge} but replaces  $B_f$ by $\hat q_{\tilde f}$, meaning \eqref{eq:mb_interval0} is replaced by:
\begin{equation}
    \label{eq:mb_interval}
    \begin{aligned}
    \varphi_1[n+1] := \\
    \mathop{\text{argmax}}_{k} \left \{ |V_{\tilde f}^g[n+1,k]|,  |k - 
    \varphi_1[n] - \hat q_{\tilde f}[n,\varphi_1[n]]\frac{N}{L^2}  |  \leq  C \right \},
\end{aligned}
\end{equation}
the user-defined constant $C$ compensating for potential local frequency modulation estimation errors. 

MB-RD is proved to be slightly sensitive to $C$ when the noise level is low \cite{colominas2020fully} (the simulations in that paper 
only considered a SNR larger than 0 dB), but the robustness of  $\hat q_{\tilde f}$ to heavier noise needs to be further investigated to fully validate this technique.
Furthermore, though MB-RD is more adaptive than S-RD, both techniques are based on the assumption that the IF of a mode at each time index can be associated with a LMMF, 
which may not be  the case in heavy noise situations. Another limitation of S-RD and MB-RD is that they build the ridges  
one after the other using the peeling algorithm recalled in Section  \ref{sec:ridge}:  if the ridge detection fails for one mode,  it will also fail for the next ones.   
To deal with all these issues, the concept of \emph{relevant ridge portions} (RRPs) is introduced in the following section, and subsequently 
used to define a new RD not based on the just mentioned peeling algorithm. 

\section{Definition of a New Robust Ridge Detector}
\label{sec:defRD} 
\subsection{Definition of Relevant Ridge Portions}
\label{sec:RRPs}
One limitation of MB-RD is that it assumes the modulation operator $\hat q_{\tilde f}$  is accurate in noisy situations which is not necessarily the case. 
Another limitation is that it cannot deal with situations where the IF of a mode is not associated with a continuous chain of LMMFs as illustrated in Fig. \ref{Fig1}. 
Therefore, to try to build a continuous chain of LMMFs based on $\hat q_{\tilde f}$ is not relevant in these situations, and to try to associate with the IF of a mode  
a set of \emph{ridge portions} (RPs) seems to be more to the point.

From now on, $[n,m[n]]$ denotes a generic LMMF, $m[n]$ being a frequency index, namely $[n,m[n]]$ is one of the LMMFs at time index $n$. 
To keep the adaptivity of MB-RD while taking into account potential inaccuracies in the modulation estimation given by $\hat q_{\tilde f}$, we define RP by linking LMMFs at which 
the value of  $\hat q_{\tilde f}$ corresponds to a stable orientation.  For that purpose we introduce the following: 
\begin{definition}
\label{defi1}
Let  $[n,m[n]]$ and $[n+1,m[n+1]]$ be two LMMFs, then define:  
{
\small
\begin{eqnarray}
 \left ( [n,m[n]] \sim [n+1,m[n+1]] \right )  \Leftrightarrow \hspace{1 cm}  \nonumber\\
 \left \{ \begin{array}{l}
m[n+1] := 
\mathop{\text{argmin}}\limits_{k} \left \{  |k- m[n] - \hat q_{\tilde f}[n,m[n]] \frac{N}{L^2}|,  \right . \\
                 \left . \hspace{4  cm} \textrm{s.t. }  [n+1,k] \textrm{ LMMF}\right \} \\
m[n] := 
 \mathop{\text{argmin}}\limits_{k} \left \{ |m[n+1] - k - \hat q_{\tilde f}[n+1,m[n+1]] \frac{N}{L^2}|, \right . \\
                \left .  \hspace{4  cm}  \textrm{s.t. } [n,k] \textrm{ LMMF}\right \}.
\end{array} 
\right .
\end{eqnarray}
}
\end{definition} 
Definition \ref{defi1} tells us that $[n+1,m[n+1]]$ (resp. $[n,m[n]]$) is the closest LMMF to $[n,,m[n]]$ (resp. $[n+1,m[n+1]]$) at time index $n+1$ (resp. $n$) in the direction given by 
$\hat q_{\tilde f}[n,m[n]]$ (resp. $-\hat q_{\tilde f}[n+1,m[n+1]]$). 
So $[n,m[n]] \sim [n+1,m[n+1]]$ also means that  $\hat q_{\tilde f}$ computed at these LMMFs corresponds to a stable orientation. 

Relation $\sim$ can then be used to define RPs by extending relation $\sim$ to LMMFs associated with time indices $n$ and $n_0$, such that $n \geq n_0+1$, by introducing: 
\begin{definition}
\label{defi2} 
Let $[n,m[n]]$ and  $[n_0,m[n_0]]$ be two LMMFs such that $n \geq n_0+1$, then define:
\begin{eqnarray}
\left ( [n_0,m[n_0]] \leftrightarrow  [n,m[n]] \right )
\Leftrightarrow  \nonumber \\
\begin{cases}
\exists \ [n_0+1,m[n_0+1]],..., [n-1,m[n-1]]   \textrm{  LMMFs} \\
\forall i =n_0,\cdots,n-1, \ [i,m[i]] \sim [i+1,m[i+1]] 
\end{cases}
\end{eqnarray}
\end{definition} 

\noindent A RP ${\cal R}$ containing LMMF $[n_0,m[n_0]]$,  is finally defined  by: 
\begin{equation}
\label{def:rRRP}
{\cal R}[n_0,m[n_0]] = \left \{   [n,m[n]],  \textrm{s.t. } [n,m[n]] \leftrightarrow [n_0,m[n_0]] \right \}.
\end{equation}
Note that, if $ [n,m[n]] \in {\cal R}[n_0,m[n_0]] $ then  ${\cal R}[n,m[n]] =  {\cal R}[n_0,m[n_0]]$. 

The definition of RPs does not take into account the magnitude of the STFT at LMMFs, 
and thus some RPs may correspond to noise. We now explain how to suppress most 
RPs related to noise. For that purpose, let us assume that the added complex noise $\varepsilon$ (see Eq. \eqref{def:ftilde}) is Gaussian white with variance 
$\sigma_\varepsilon^2$. Then it can be shown that  $V_{\varepsilon}^g[n,k]$ is also Gaussian with zero mean and satisfies \cite{Pham2018a}: 
\begin{eqnarray*}
\textrm{Var} \left ( \Re \{ V_{\varepsilon}^g [n,k]  \} \right ) =  \textrm{Var} \left ( \Im \{ V_{\varepsilon}^g [n,k]  \} \right ) = \sigma_\varepsilon^2 \| g \|^2_2,
\end{eqnarray*}
where $\Im\{X\}$ is the imaginary part of complex number $X$.  Then, remarking that $\frac{|V_{\varepsilon}^g|^2}{ \sigma_\varepsilon^2 \| g \|^2_2}$ 
is $\chi_2$ distributed with two degrees of freedom and assuming the variance of the noise $\sigma_\varepsilon^2$ is known, 
the probability that  $|V_{\varepsilon}^g [n,k]| \geq  \beta \sigma_\varepsilon  \|g\|_{2}$ is smaller than $10\%$ and $0.01\%$ if $\beta = 2$ and $3$,  respectively.
So by considering only the LMMFs $[n,m[n]]$ such that  $|V_{\tilde f}^{g} [n,m[n]]| \geq  \beta \sigma_\varepsilon  \|g\|_{2}$ with $\beta \geq 2$ one 
removes many LMMFs corresponding to noise. Note that, in this paper, to estimate   
$\gamma = \sigma_\varepsilon  \|g\|_{2}$, the robust estimator proposed in \cite{Donoho1994}:  
\begin{align*}
\label{signalstdnoise}
\hat \gamma = \frac{ \mathop{\textrm{median}}  \left|\Re \left \{ V_{\tilde f}^g[n,k] \right \}_{n,k} \right|}{0.6745},
\end{align*} 
is used, in which  $\textrm{median}$ represents the median of the coefficients. 
Based on this analysis, one defines 
\begin{equation}
 \label{def:S}
 {\cal S} (\beta) = \left \{ [n,k] ,  |V_{\tilde f}^g [n,k]| \geq \beta \hat \gamma \right \},
 \end{equation}
and then, to eliminate most of the RPs corresponding to noise, one slightly modifies Definition \ref{defi2} into 
\begin{definition}
\label{defi3} 
Let $[n,m[n]]$ and  $[n_0,m[n_0]]$ be two LMMFs such that $n \geq n_0+1$, then define:
{\small 
\begin{eqnarray}
[n_0,m[n_0]] \leftrightsquigarrow  [n,m[n]]  
\Leftrightarrow \nonumber \\  
\begin{cases}
\exists \ [n_0+1,m[n_0+1]],\cdots, [n-1,m[n-1]]   \textrm{  LMMFs } \\
\forall i =n_0,\cdots,n-1, \  \begin{cases}
                                          [i,m[i]] \sim [i+1,m[i+1]]\\ 
                                          [i,m[i]] \in  {\cal S} (\beta)
                                         \end{cases}
\end{cases}
\end{eqnarray}
}
\end{definition} 
The LMMFs connected by means of Definition \ref{defi3} are called from now on \emph{relevant ridge portions} (RRPs), 
and a LMMF belonging to a RRP is called a \emph{relevant LMMF}.

A crucial issue is then how to fix the parameter $\beta$.  From now on, we consider that $g$ is a Gaussian window, first because it has 
the advantage that the windows $g'$, $tg$, $g''$ and $tg'$ have analytical expressions, $g'$ being even proportional to $tg$. So,  
for the computation of $\hat q_{\tilde f}$ used in the definition of RRPs, only 3 different STFTs are needed. 
The second advantage is that the expression of the STFT of a signal that can be locally approximated by a linear chirp is particularly simple \cite{behera2018theoretical}. 
Indeed, let  $g$ be the discrete Gaussian window $g[n] = e^{-\pi \frac{n^2}{\sigma^2 L^2}}$, 
where $\sigma $ is associated with the STFT whose modulus minimizes the R\'enyi entropy. Such a choice for $\sigma$  
is proved to minimize interferences between the modes in the TF plane \cite{baraniuk2001measuring,meignen2020use}. 
With such a window, if $f$ can be locally approximated by a linear chirp with constant amplitude $A$, i.e. $f[n] \approx A e^{2i \pi \phi[n]}$, with $\phi$ a 
second order polynomial,  
one has \cite{behera2018theoretical}:
\begin{equation}
\label{eq:STFT_linear_chirp}
\begin{aligned}
|V_{f}^g[n,k]| \approx A L \sigma (1+\sigma^4 \phi''[n]^2)^{-\frac{1}{4}} 
e^{-\pi \frac{\sigma^2(k\frac{L}{N}-\phi'[n])^2}{1+\sigma^4\phi''[n]^2}},
\end{aligned}
\end{equation}
whose standard deviation is:
\begin{equation}
\label{def:std-LC}
\begin{aligned}
\Delta_{LC}[n] = \frac{1}{\sqrt{2 \pi} \sigma} \sqrt{1+ \sigma^4 \phi''[n]^2} .
\end{aligned}
\end{equation}
In that case, the LMMF associated with the largest STFT modulus maximum at time index $n$ has its ordinate in the interval 
${\cal I} [n] = [(\phi'[n] - \Delta_{LC}[n])\frac{N}{L}, (\phi'[n] + \Delta_{LC}[n]) \frac{N}{L}]$.  When some noise is added, we assess the proportion corresponding to the number of relevant LMMFs with ordinate in ${\cal I}[n]$ when $n$ varies,
 namely:
\begin{equation}
\label{def:P}
{\cal P}(\beta)=\frac{\# \left \{ [n,m[n]]   \in  {\cal S} (\beta) \textrm{ with } m[n] \in {\cal I}[n] \right \}}{L},
\end{equation}  
where $\# X$ denotes the cardinal of the set $X$. 
Our motivation for the choice for $\beta$ is to remove as much noise as possible while keeping ${\cal P} (\beta)$ large enough so that a relevant LMMF has 
its ordinate in ${\cal I}[n]$ for most $n$, because RRPs, on which we are going to found our new RD, will depend  on relevant LMMFs. 
Note that our choice for $\beta$ is also driven by heavy noise situations (typically input SNR = -10 dB). To find an appropriate value for $\beta$, we  
compare ${\cal P}(\beta)$ with the proportion of the number of time indices $n$, at which the relevant LMMF corresponding to the largest STFT 
modulus maximum has its ordinate in ${\cal I}[n]$.   

\begin{figure}[!htb]
	
	\centering
	\begin{minipage}{0.48 \linewidth}
		\begin{tabular}{c}
		\includegraphics[width = 4 cm, height = 1.2 cm] {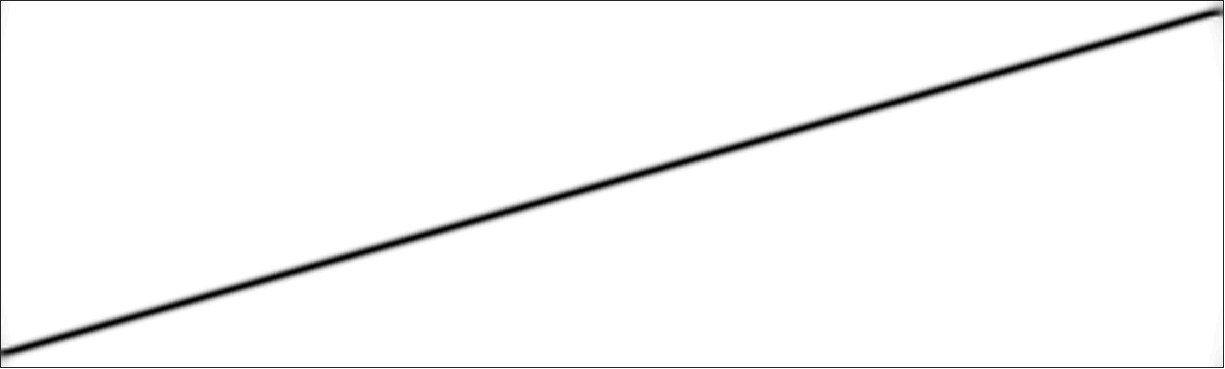}\\
		\includegraphics[width = 4 cm, height = 1.2 cm] {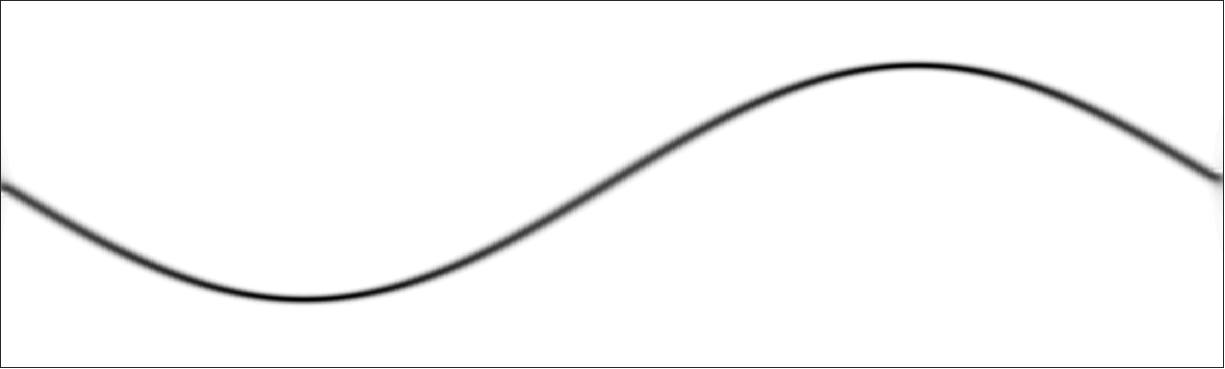}\\
		\includegraphics[width = 4 cm, height = 1.2 cm] {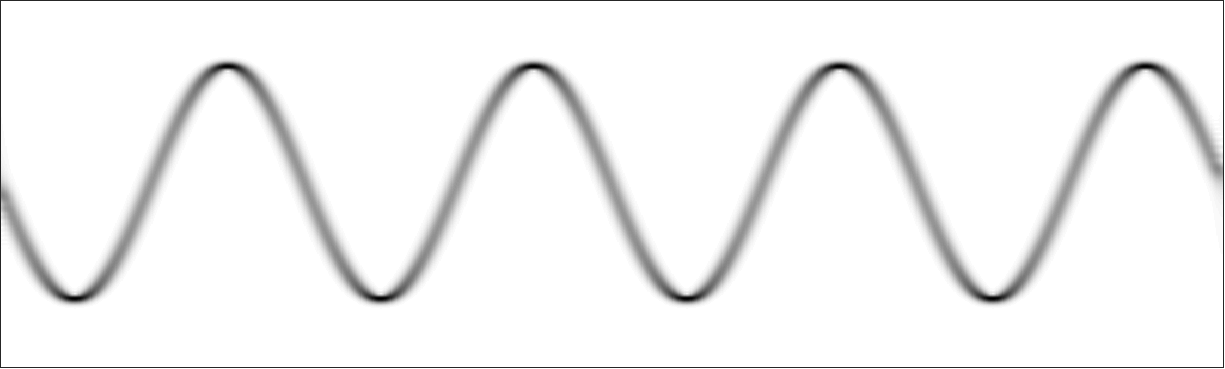}\\
		\end{tabular}\\
		\vspace{1mm}\\
		\centering
	\end{minipage}
	\begin{minipage}{0.48 \linewidth}
		\includegraphics[width= 6 cm, height = 4 cm] {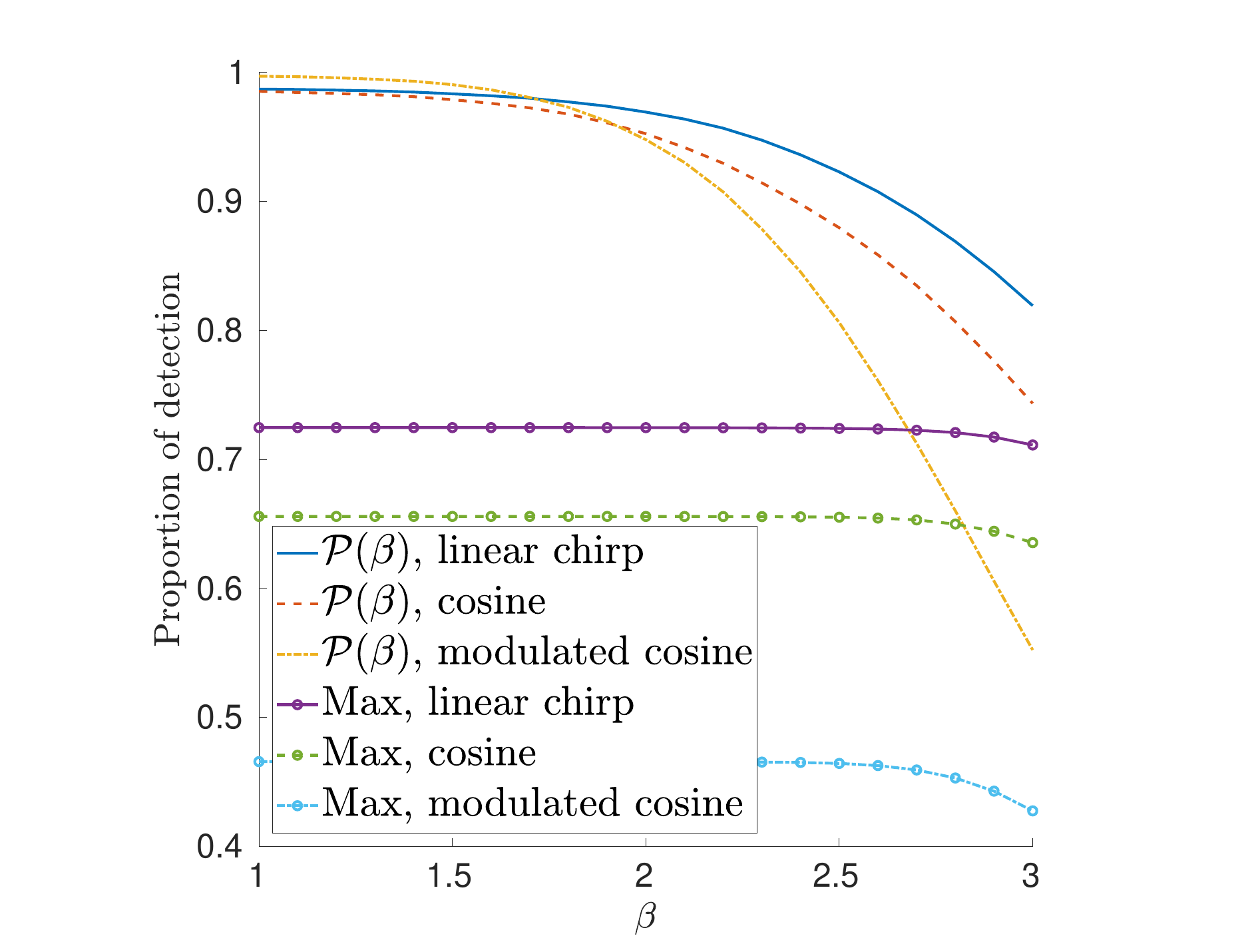}\\
    \end{minipage}\\
	{\hspace{3.5cm} (a) \hspace{8cm} (b) \hfill}
	
\caption{(a): from top to bottom: STFT of a linear chirp, of a signal with cosine phase, and of a signal with oscillatory phase; (b): ${\cal P}(\beta)$ for these three signals when the input SNR equals -10 dB  as well as the proportion of time indices $n$, at which the relevant LMMF with ordinate in ${\cal I}[n]$ corresponds to the largest STFT modulus maximum  
(curves denoted by "Max, linear chirp",  "Max, cosine", and "Max, modulated cosine"). The results are average over 100 noise realizations.}
\label{Fig2}
\end{figure}

We carry out such a study for the three signals of Fig. \ref{Fig2} (a), which from top to bottom are more and more modulated,
 and get the results of Fig. \ref{Fig2} (b) corresponding to an input SNR of -10 dB. 
 First, we note that the more modulated a signal is, 
 the fewer relevant LMMFs are located in the region of interest when $\beta$ increases.  
These simulations also highlight the fact that $\beta = 2$ is appropriate.
Indeed, ${\cal P}(3)$ is too low, meaning too many LMMFs in the vicinity of the true IF location are discarded, and, 
if  $\beta \in [1,2]$, ${\cal P}(\beta)$ is almost constant for all the signals and then decay faster for larger $\beta$ when the modulation is higher.
So to take $\beta =2$ is a good trade-off between a high value for  ${\cal P}(\beta)$ and small probability of false 
detection. From now on, $\beta$ equals $2$ unless mentioned otherwise.

\subsection{Grouping RRPs Based on Basins of Attraction }  
\label{sec:basins} 
We are now going to explain how to group RRPs based on the \emph{basins of attraction} (BAs) associated with RRPs. 
The definition of BAs is tightly connected with the notion of TF reassignment, whose goal is to relocalize the energy of the STFT at TF location 
$(\frac{n}{L}, k\frac{L}{N})$ to a meaningful 
location $(\hat \tau  [n,k],\hat \omega [n,k])$, defined by  \cite{Auger1995}
 \begin{equation}   
 \label{def:reassign_op}  
 \begin{aligned}
  \hat \omega [n,k] &= k\frac{L}{N} - \Re \left \{ \frac{1}{2i\pi} \frac{V_f^{g'} [n,k]}{V_f^{g}[n,k]}   \right \} \\
  \hat \tau  [n,k] &= \frac{n}{L} + \Re \left \{ \frac{V_f^{tg} [n,k]}{V_f^{g}[n,k]} \right \}.
  \end{aligned}
  \end{equation}  

  In practice, each point in the TF plane corresponding to a non-zero STFT is reassigned to a LMMF, and conversely, each LMMF can be associated with  a 
  region of the TF plane by means of this reassignment process. 
  The set of points that are reassigned to one of the relevant LMMF of a RRP, is called the  BA of that RRP. 
  It is worth noting here that the concept of BAs has already been used for the purpose of mode retrieval in  \cite{Meignen2016a}, and we use the 
  same procedure to build them. 
  
  In a noisy context, we propose to consider the set of BAs associated with RRPs containing at least one LMMF $[n,m[n]]$ in  ${\cal S} (3)$, 
  and, in these BAs, to keep only the points $[n,k]$ in ${\cal S} (2)$. With this two threshold procedure  
  we only keep the BAs that most probably correspond to the signal, and then, in these BAs, we only keep the points at which the STFT is above the 
  noise level with great confidence.  
  More formally, let ${\cal B}_i$ be the BA corresponding to RRP ${\cal R}_i$, the set of points we actually consider in  ${\cal B}_i$ corresponds to :
  \begin{equation}
  \label{def:basinthresh}
  {\cal B}_i^{HT} = \left \{\begin{array} {c c} 
  {\cal B}_i \bigcap {\cal S} (2) & \textrm{ if }  {\cal R}_i  \bigcap  {\cal S} (3) \neq \emptyset\\
  \emptyset & \textrm{ otherwise },
  \end{array}
  \right .
  \end{equation}
  the superscript $HT$ standing for \emph{hard-thresholding}. 
  We then gather together connected ${\cal B}_i^{HT}$s in the TF plane to obtain a set of larger TF regions which is  
  denoted by $\{ ({\cal C}_j^{HT})_j\}$ in the sequel. 
\subsection{New RD Definition} 
\label{sec:newrd1}
To define the new RD, assuming the number $P$ of modes is known,  we use the set $\{({\cal C}_j^{HT})_j\}$, the elements of which correspond to specific sets of time indices. 
We first select the $P$ elements in that set that coexist on the longest set of time indices, since these elements are very likely to belong to different modes.        
We denote by $({\cal C}_p^0)_{p=1,\cdots,P}$ these $P$ elements supposed to be reordered according 
to increasing frequencies. Then, to perform a first approximation of the ridges, we consider a spline approximation based on the LMMFs in 
$ {\cal A}_p^0= {\cal C}_p^0 \bigcap {\cal S}(3)$, as follows: 
\begin{eqnarray}
\label{def:least_square}
s_{p}^0 =
\mathop{\textrm{argmin}}_{s}  \nonumber \\
 \left [ 
(1-\lambda)\sum_{[n,m[n]] \in {\cal A}_{p}^0}
                                                        |m[n] \frac{L}{N} - s(\frac{n}{L})|^2 |V_{\tilde f}^g[n,m[n]]| \right . \nonumber \\
\left .  + \lambda \int_0^1 (s''(t))^2 dt, \right  ],
\end{eqnarray}
$p = 1,\cdots,P$, where $s$ is a cubic spline and $\lambda$ a user defined parameter.  
Then, if $s_{p}^0$ intersects ${\cal C}_j^{HT}$ not in $({\cal C}_p^0)_{p=1,\cdots,P}$,  ${\cal C}_j^{HT}$ is added to that set 
and the minimization \eqref{def:least_square} is recomputed with the updated set. 
Such a procedure is iterated until no new elements in $\{ ({\cal C}_j^{HT})_j\}$ are intersected by the updated $s_p^0$. Note that this procedure computes simultaneously the 
approximations for the $P$ ridges. For the sake of simplicity, we still denote by $(s_p^0)_{p=1,\cdots,P}$ the set of cubic splines obtained at the end of this procedure, and by  
$({\cal C}_p^0)_{p=1,\cdots,P}$ the regions involved in the minimization process.  After this step, we define an energy associated with the spline approximation as follows:  
\begin{figure*}[!htb]
\centering
\begin{minipage}{0.32\linewidth}
	\begin{tabular}{c}
	\includegraphics[width=\textwidth,height = 4 cm] {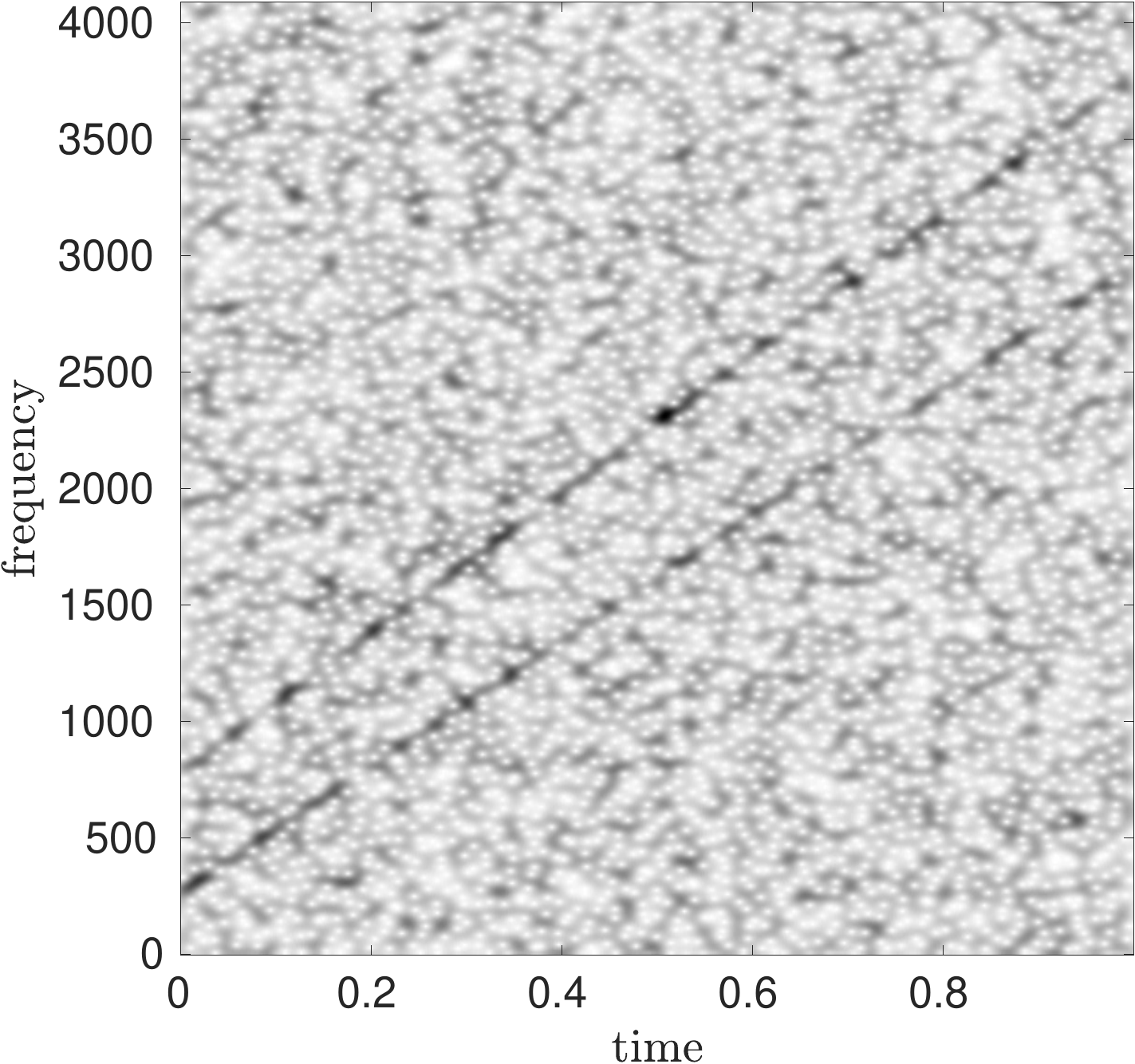}\\
	\hspace{0.6 cm} (a)
	\end{tabular}
\end{minipage}
\begin{minipage}{0.32\linewidth}
    \begin{tabular}{c}
	\includegraphics[width=\textwidth,height = 4 cm] {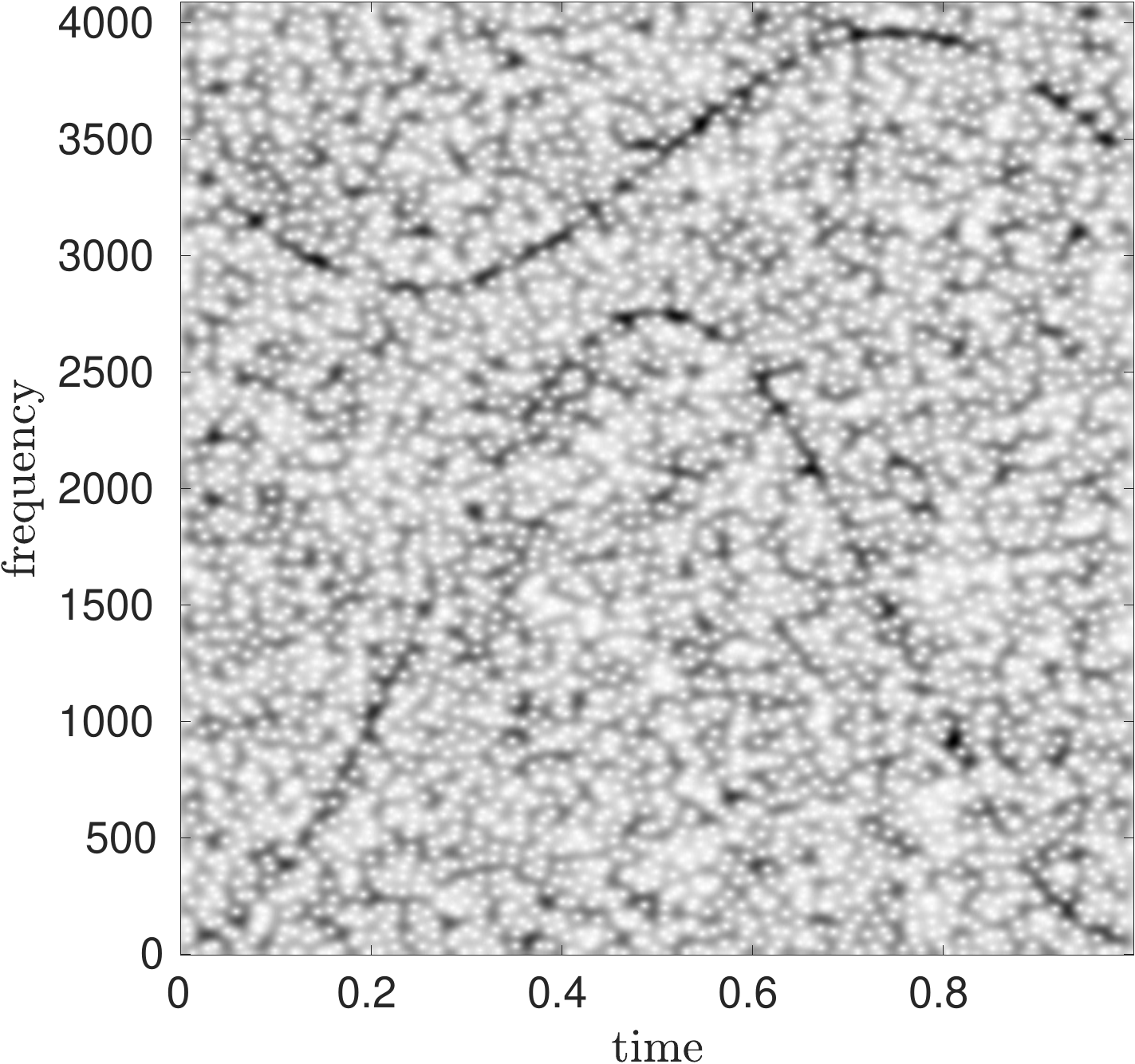}\\
    \hspace{0.6 cm} (b)
    \end{tabular}
\end{minipage}
\begin{minipage}{0.32\linewidth}
    \begin{tabular}{c}
	\includegraphics[width=\textwidth,height = 4 cm] {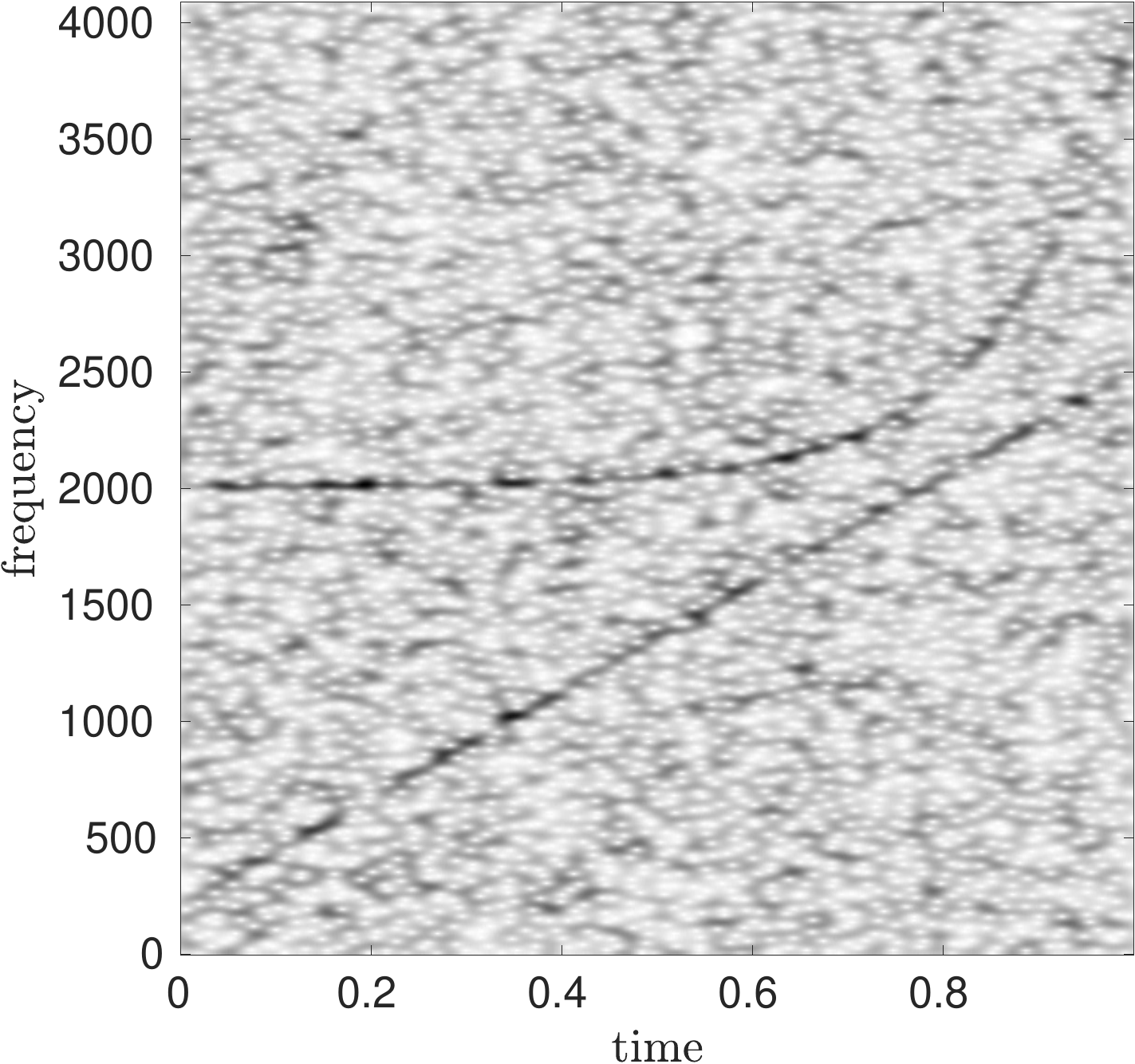}\\
    \hspace{0.6 cm} (c)
    \end{tabular}
\end{minipage}   
\centering
\begin{minipage}{0.32\linewidth}
	\begin{tabular}{c}
	\includegraphics[width=\textwidth,height = 4 cm] {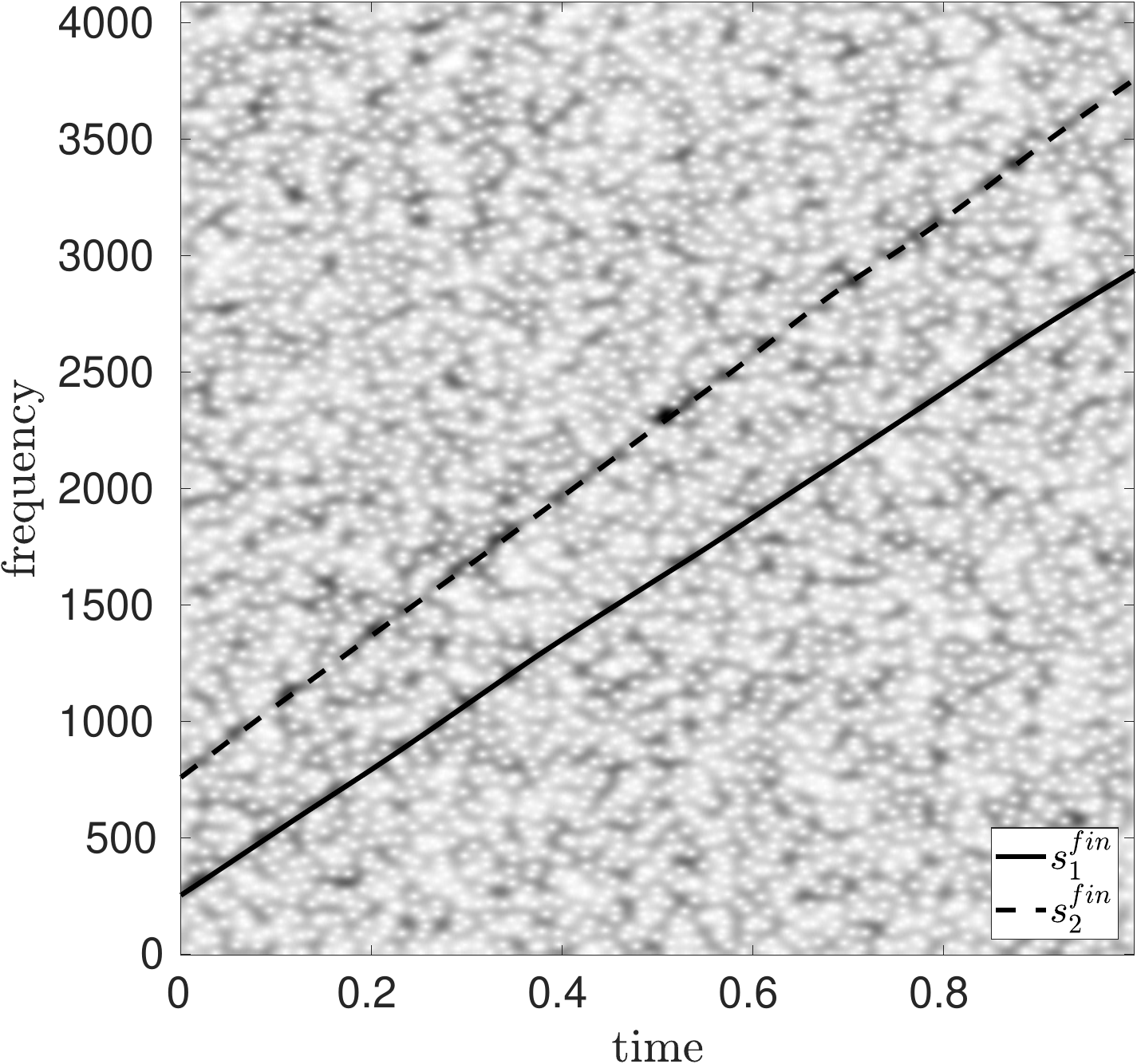}\\
	\hspace{0.6 cm} (d)
	\end{tabular}
\end{minipage}%
\begin{minipage}{0.32\linewidth}
	\begin{tabular}{c}
	\includegraphics[width=\textwidth,height = 4 cm] {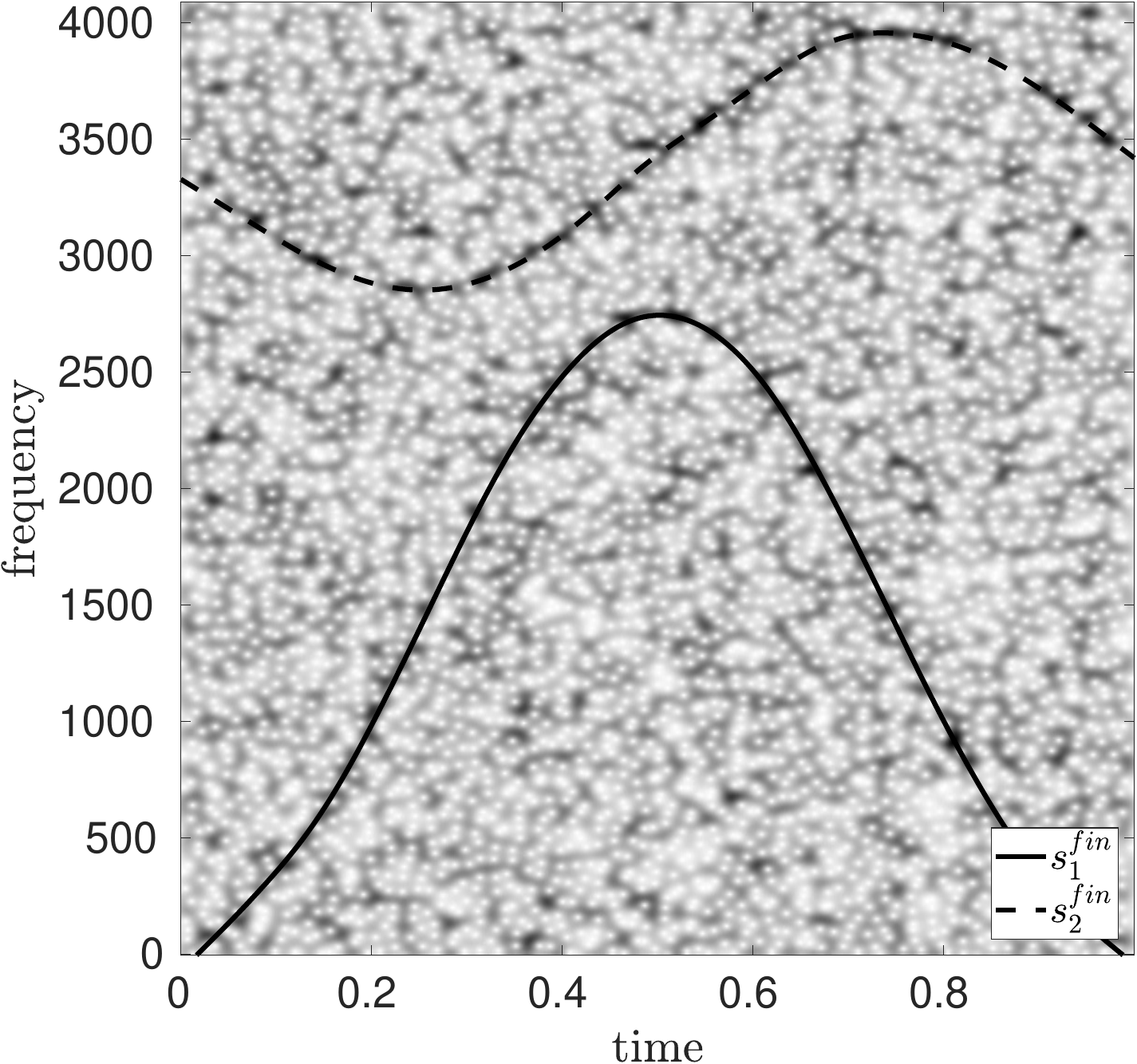}\\
    \hspace{0.6 cm} (e)
    \end{tabular}
\end{minipage}
\begin{minipage}{0.32\linewidth}
   \begin{tabular}{c}
	\includegraphics[width=\textwidth,height = 4 cm] {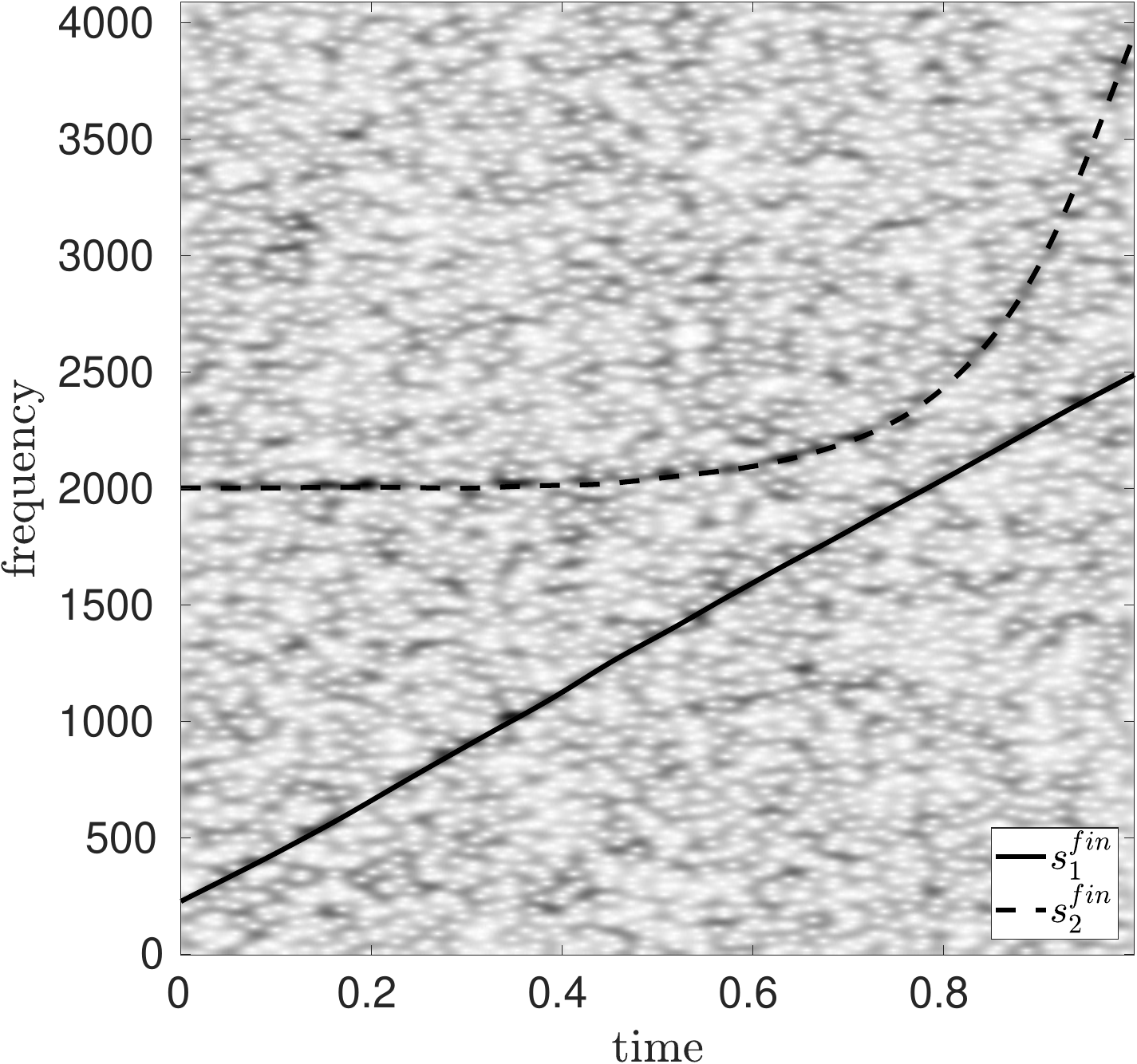}\\
    \hspace{0.6 cm} (f)
   \end{tabular}
\end{minipage}
\caption{(a): STFT modulus of two noisy linear chirps (SNR = -10 dB, $\sigma =0.0188$); 
(b): STFT modulus of two noisy modes with cosine phase, with different modulation (SNR = -10 dB, $\sigma =0.0175$); 
(c): STFT modulus of a signal made of a linear chirp and a mode with exponential phase (SNR = -10 dB, $\sigma =0.0241$); 
(d): $(s_p^{fin})_{p=1,2}$ computed for the signal displayed in (a);
(e): same as (d) but for the signal whose STFT modulus is displayed in (b); 
(f): same as (d) but for the signal whose STFT modulus is displayed in (c).}
\label{Fig4}
\end{figure*}
\begin{eqnarray}
\label{def:energies}
E^0_p :=  \sum_{
[n,m[n]] \in {\cal A}_{p}^0, 
[n, \lfloor s_p^0 [n] \frac{N}{L} \rceil ] \in  {\cal C}_p^{0}
}  
 |V_{\tilde f}^g [n,m[n]]|, 
\end{eqnarray}
in which $\lfloor X \rceil$ denotes the nearest integer to $X$.
Note that the condition $[n, \lfloor s_p^0 [n] \frac{N}{L} \rceil ] \in  {\cal C}_p^{0}$ is added so as to assign some energy only when the approximating spline 
is close to the LMMFs used for its computation.

The set $(s_p^0)_{p=1,\cdots,P}$ consists of a first approximation for the $P$ ridges. 
Then, one considers the set $({\cal C}_p^1)_{p=1,\cdots,P}$  of $\{({\cal C}_j^{HT})_j\}$ that coexist on the second longest set of time indices, and solve the new following 
optimization problem, putting ${\cal A}_{p}^1 = ({\cal C}_{p}^0 \bigcup {\cal C}_{p}^1) \bigcap {\cal S}(3)$ :  
\begin{eqnarray}
\label{def:least_square0}
s_{p}^1 = \mathop{\textrm{argmin}}_{s} \nonumber \\ 
 \left [ (1-\lambda) \sum_{
 [n,m[n]] \in {\cal A}_{p}^1} | m[n]  \frac{L}{N} - s(\frac{n}{L}) |^2 | V_{\tilde f}^g[n,m[n]] | \right . \nonumber \\
\left .
+ \lambda \int_0^1 (s''(t))^2 dt, \right ],
\end{eqnarray}
 $p = 1,\cdots,P$. If $s_{p}^1$ intersects ${\cal C}_{j}^{HT}$ not in ${\cal C}_{p}^0 \bigcup {\cal C}_{p}^1$, the former is added to the latter, and the minimization 
\eqref{def:least_square0} recomputed. This process is iterated until no new elements in $\{({\cal C}_j^{HT})_j\}$ are intersected. 
We still denote by  $(s_p^1)_{p=1,\cdots,P}$ the set of splines obtained at the end of this procedure, and we associate an energy $E^{1}_p$ to the spline $s_{p}^1$  
the same way as in \eqref{def:energies}. 

Such a procedure is iterated until it is not possible to find a new set of $P$ elements in $\left \{ ({\cal C}_j^{HT})_j \right \}$ coexisting for some time indices. 
At the end of this procedure, we keep the set of $P$ splines associated with the largest energies and that do not intersect.
In the sequel, we denote by $(s_p^{fin})_{p=1,\cdots,P}$ this set and by RRP-RD this new ridge detector. 

We display on the first row of Fig. \ref{Fig4} the modulus of the STFT of noisy two-mode signals 
made either of two linear chirps, of two modes with cosine phase, or  
of a linear chirp plus an exponential chirp. 
In each case, we consider complex white Gaussian noise and the input SNR equals $-10$ dB. 
On the second row of Fig. \ref{Fig4}, we display $(s^{fin}_p)_{p=1,2}$ computed with the optimization procedure just described.
We notice that RRP-RD seems to be well adapted to deal with MCSs in the presence of heavy noise 
regardless of the modulation of the modes. 
\subsection{Analysis of the Computational Cost}
\label{sec:comput}
We can analyze the computational cost of RRP-RD by considering each of the above three steps separately. 
The first step, consisting of the definition of RRPs is achieved in $O(N_{LMMF})$ operations, where $N_{LMMF}$ denotes the number of LMMFs (for each LMMF,  the computational cost is related to the establishment of connections with neighboring LMMFs based on  $\hat q_{\tilde f}$). The second step of the algorithm consisting 
of the definition of the basins of attraction is carried out in  $O(LN)$ operations (the computational cost corresponds to the reassignment  of TF coefficients to the closest RRP). 
Finally, the computational cost of the third step is mostly related to the initialization of the weighted spline approximation, 
namely the computation of the set $({\cal A}_p^0)_{p=1,\cdots,P}$ (the computational cost is linear with respect to the number of regions in $\{({\cal C}_j^{HT})_j\}$). 
 
\section{Estimation of the Number of Modes}
\label{sec:nummodes}
To compute an estimation of the number of modes, we  first introduce the energy of ${\cal R}_i$ as: 
  \begin{equation}
  \label{def:energy_basins} 
  E({\cal R}_{i}) = \sum_{[n,k] \in {\cal R}_{i} \bigcap {\cal S} (3)}  |V_{\tilde f}^g[n,k]|,
  \end{equation}
   and that of ${\cal C}_j^{HT}$ by:
  \begin{equation}
  \label{def:energy_regions} 
  E({\cal C}_j^{HT}) = \sum_{i, {\cal R}_{i} \subset {\cal C}_j^{HT}} E ({\cal R}_{i}).
  \end{equation} 
 Finally, for any $[n,m[n]]$ belonging to $ {\cal C}_j^{HT}$ for some $j$, we set $E[n,m[n]] = E({\cal C}_j^{HT})$ (in any other circumstances $E[n,k]$ is set to $0$).

\begin{figure}[!htb]
\centering
\includegraphics[width=5 cm, height = 5 cm] {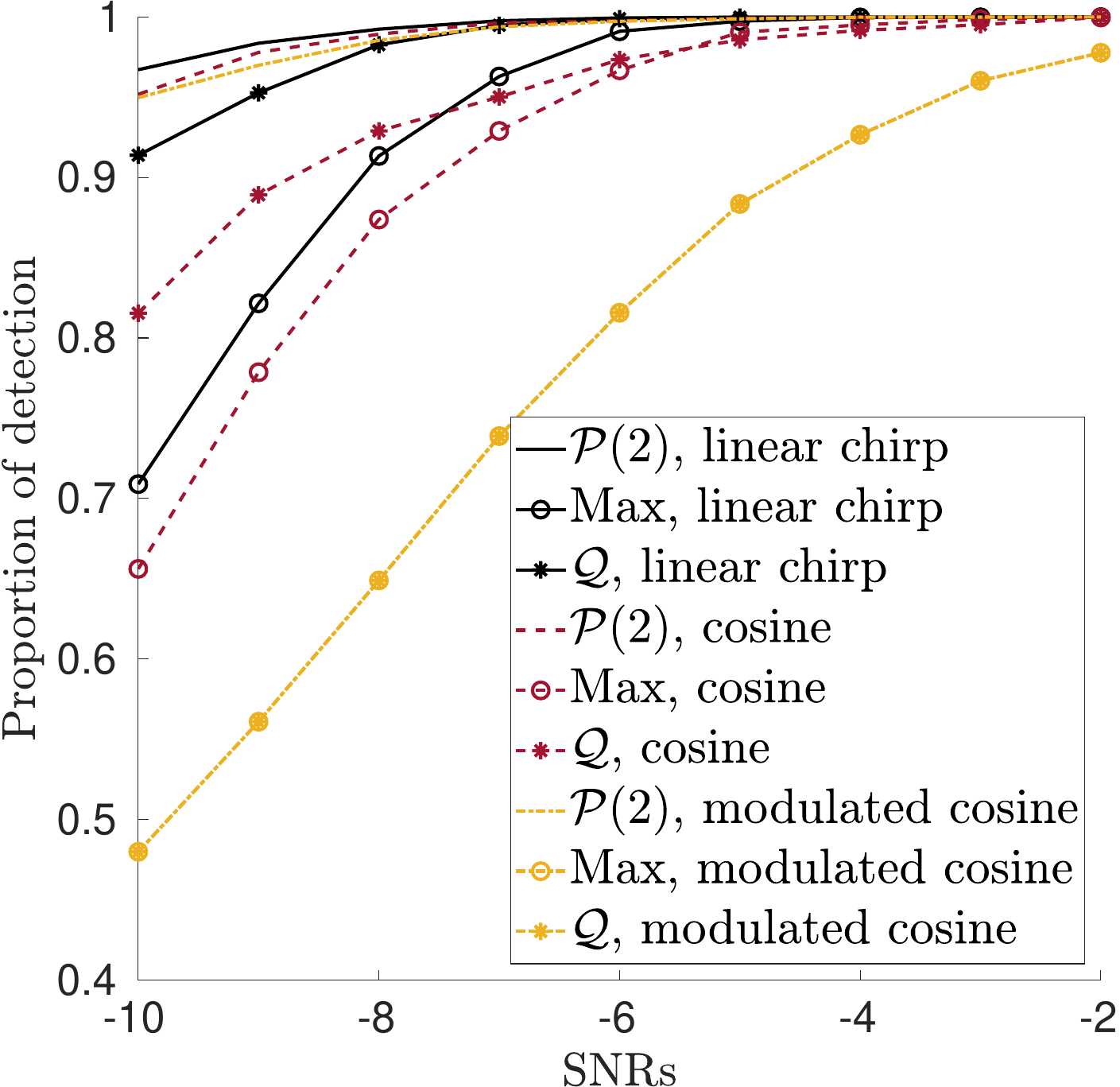}\\
\caption{${\cal P}(2)$, ${\cal Q}$ and the proportion of time indices $n$ at which the largest STFT modulus maximum located is located in ${\cal I}[n]$ 
(still with $\beta=2$, denoted by "Max, cosine ","Max, linear chirp", and "Max, modulated cosine "), when the input SNR varies and for the three signals of Fig. \ref{Fig2} (a). 
The results are averaged of 30 noise realizations.}
\label{Fig3}
\end{figure}

We already noticed that for a monocomponent signal and  at high noise level, a relevant LMMF with ordinate in ${\cal I}[n]$ may not correspond to the  
largest STFT modulus maximum at that time index. But, such a LMMF most probably corresponds to the global maximum of $E$ at that time.
To prove this, let us introduce: 
\begin{equation}
\label{def:Q}
{\cal Q}=\frac{\# \left \{ n, \max\limits_k  (E[n,k]) > 0 \textrm{ and } (\mathop{\textrm{argmax}}\limits_k  E[n,k])   \in {\cal I}[n]  \right \}}{L},
\end{equation} 
and then compute, for the three signals of Fig. \ref{Fig2} (a) and when the input SNR varies,  ${\cal Q}$, ${\cal P}(2)$ and the proportion of time indices 
$n$ at which a relevant LMMF with ordinate in ${\cal I}[n]$ corresponds to the largest STFT modulus maximum. 
The results depicted in Fig. \ref{Fig3} show that ${\cal Q}$ is very similar to ${\cal P}(2)$ for a linear chirp, 
meaning a LMMF with ordinate in ${\cal I}[n]$ at time index $n$ is very likely to correspond to the maximum of $E$ at that time. 
For the second signal of Fig. \ref{Fig2} (a), ${\cal Q}$ is significantly smaller than  ${\cal P}(2)$ but still much larger than 
the proportion of time indices $n$ at which the frequency corresponding to the largest STFT modulus maximum is located in ${\cal I}[n]$. 
If  the signal has a more oscillating phase, as the third signal of Fig. \ref{Fig2} (a), the global maxima of $E$ correspond to global maxima of STFT moduli. 
From this study, it transpires that  $E$ evaluated at LMMFs better reflects the presence of a mode than STFT modulus at these locations. This is why we are now going 
to use $E$ to estimate the number of modes.

Let us first define the global energy of the decomposition as:
\begin{equation}
\label{eq:gobal}
E_{glob} = \sum_{n=0}^{L-1} \sum_{k=0}^{N-1}  E[n,k].
\end{equation} 
Then, for each time index $n$, we introduce the set of frequencies corresponding to the $P$ most energetic LMMFs as  
$
{\cal M}_{P}[n] = \left \{k, \textrm{ s.t. } E[n,k] > 0 \textrm{ is one of the }P \textrm{ largest values w. r. to } k  \right \},
$   
and define the proportion of the energy associated with these sets by: 
\begin{equation}
\label{eq:P_max_n}
E_{modes}(P) = \frac{\sum\limits_{n=0}^{L-1} \sum\limits_{k \in {\cal M}_{P}[n]}  E[n,k]}{E_{glob}}.
\end{equation}
We expect that if $P$ is smaller than the actual number of modes then $E_{modes}(P)$ is increasing  and much smaller than $1$, and that $E_{modes}(P)$ stabilizes 
when $P$ is larger than the actual number of modes. Setting $E_{modes}(0) = 0$,  we thus propose to estimate the number of modes as:
\begin{eqnarray} 
\label{def:hatP}
\begin{aligned}
\hat P = \\
 {\mathop{\textrm{argmax}}_{P}}  \ 2 E_{modes}(P)-E_ {modes}(P-1) - E_{modes}(P+1) 
\end{aligned}
\end{eqnarray}
The validity of such an estimate will be discussed in Section \ref{sec:res}.

\section{Application to Mode Retrieval}
\label{sec:mode_rec}

We now recall different mode retrieval techniques based on ridge detection, which we will compare in Section \ref{sec:res}, mainly to highlight the fact 
that a good ridge detector, though essential for mode retrieval, is not sufficient in very noisy situations. In these circumstances, we will see that other assumptions have to 
be made on the modes to enable good reconstruction. 

A first simple strategy consists of summing the coefficients above the noise level in the vicinity of the ridges computed by one of the RDs. 
For that purpose, one introduces  intervals  ${\cal I}_{p}[n]:= [{\cal I}_{p}^-[n],{\cal I}_{p}^+[n]]$, with  
\begin{equation}
\label{def:Ip}
\begin{aligned}
{\cal I}_{p}^{-}[n]   &:=& \mathop{\textrm{argmax}}\limits_{k}  \left \{ k < A [n], \ [n,k] \notin {\cal S} (2) \right \}\\
{\cal I}_{p}^{+}[n]  &:=& \mathop{\textrm{argmin}}\limits_{k}  \left \{ k  >  A[n], \ [n,k] \notin {\cal S} (2) \right \},
\end{aligned}
\end{equation}
with $A[n] = \frac{N}{L} s_p^{fin} (\frac{n}{L})$ if RD is RRP-RD and $A[n] = \varphi_p[n]$ when S-RD or MB-RD are considered.
Mode reconstruction then consists of summing 
the coefficients in ${\cal I}_{p}[n]$ for each $n$, namely   
\begin{equation}
\label{def:inv_STFT_loc}
f_p [n] \approx \frac{1}{g[0]N} \sum_{k \in {\cal I}_{p}[n]} V_{\tilde{f}}^{g}[n,k].
\end{equation}
As the intervals ${\cal I}_{p}[n]$ and ${\cal I}_{p+1}[n]$ may intersect for some time index, in such instances these intervals are 
replaced by $[{\cal I}_{p}^-[n],\frac{{\cal I}_{p}^+[n]+{\cal I}_{p+1}^-[n]}{2}]$ and $[\frac{{\cal I}_{p}^+[n]+{\cal I}_{p+1}^-[n]}{2},{\cal I}_{p+1}^+[n]]$ respectively. 
These reconstruction procedures are denoted by RRP-MR, S-MR and MB-MR when RD is RRP-RD, S-RD and MB-RD, respectively. 

An alternative technique for mode reconstruction was 
recently proposed in \cite{laurent2020novel}, and aims at locally reconstructing the modes based 
on a local linear chirp approximation for the mode. In our context, the technique proposed 
in \cite{laurent2020novel} would consider $k_0 := \lfloor s^{fin}_p(\frac{n}{L}) \frac{N}{L} \rceil$, and then the following approximation 
for the STFT of $f_p$ (see \cite{laurent2020novel} for details): 
\begin{eqnarray}
\label{eq:approx_Vfp}
V_{f_p}^g [n,k] 
\approx \nonumber \\ 
V_{\tilde f}^g [n,k_0] e^{\frac{\pi\sigma^2(1 + i (s^{fin}_p)'(\frac{n}{L}) \sigma^2)}{1 + ((s^{fin}_p)'(\frac{n}{L}))^2 \sigma^4} \left [ \frac{L(k_0-k)}{N} ( \frac{L(k_0+k)}{N} - 2 s^{fin}_p (\frac{n}{L})) \right ]},
\end{eqnarray}
in which $s^{fin}_p$ and its derivative $(s^{fin}_p)'$ are estimates of $\phi_p'$ and $\phi_p''$ respectively.
If one denotes $\tilde V_{f_p}^g$ the estimation of $V_{f_p}^g$ given by \eqref{eq:approx_Vfp}, the retrieval of $f_p$ is then carried out through:
\begin{equation}
\label{eq:retrieve-Mode}
f_p[n] \approx \frac{1}{g[0]N} \sum_{k=-\frac{N}{2}}^{\frac{N}{2}-1}  \tilde  V_{f_p}^g [n,k].
\end{equation}
This technique applied to RRP-RD will be denoted by RRP-LCR-MR (LCR standing for \emph{linear chirp reconstruction}).

A very close reconstruction formula to \eqref{eq:retrieve-Mode} can be derived recalling that $V_{f}^g [n,k]$ approximates $LV_f^g(\frac{n}{L},k\frac{L}{N})$, with 
$V_f^g(t,\eta) = \int_{\mathbb{R}} f(\tau) g(\tau-t) e^{-i2\pi \eta (\tau-t) } d\tau$, and 
that a continuous version of \eqref{eq:approx_Vfp} is  \cite{laurent2020novel}
\begin{eqnarray}
\label{eq:continuous}
V_{f_p}^g (t,\eta)  \approx   \tilde V_{f_p}^g (t,\eta) = V_{\tilde f}^g (t,s_p^{fin}(t)) e^{\frac{\pi\sigma^2(\eta- s^{fin}_p(t))^2}{1 -i (s^{fin}_p)'(t) \sigma^2}}.
\end{eqnarray}
Indeed, when $\frac{L}{N}$ is small, we may write that: 
\begin{eqnarray}
\label{eq:retrieve-Mode1}
\begin{aligned}
f_p[n] & = \frac{1}{g[0]N} \sum_{k=-\frac{N}{2}}^{\frac{N}{2}-1}   V_{f_p}^g [n,k] \\
&\approx \frac{L}{g[0]N}  \sum_{k=-\frac{N}{2}}^{\frac{N}{2}-1}  V_{f_p}^g(\frac{n}{L},\frac{kL}{N}) \\
&\approx \frac{1}{g[0]}  \int_{-\frac{L}{2}}^{\frac{L}{2}}  V_{f_p}^g(\frac{n}{L},\eta) d\eta \approx \frac{1}{g[0]}  \int_{-\frac{L}{2}}^{\frac{L}{2}}  \tilde V_{f_p}^g(\frac{n}{L},\eta) d\eta \\
&\approx \frac{1}{g[0]}  \int_{\mathbb{R}} V_{\tilde f}^g(\frac{n}{L},s_p^{fin}(\frac{n}{L})) e^{-\frac{\pi \sigma^2(\eta-s_p^{fin}(\frac{n}{L}))^2}{1-i (s_p^{fin})'(\frac{n}{L}) \sigma^2}} d\eta\\
&=  V_{\tilde f}^g(\frac{n}{L},s_p^{fin}(\frac{n}{L})) \int_{\mathbb{R}}  e^{-\frac{\pi \sigma^2(\eta-s_p^{fin}(\frac{n}{L}))^2}{1-i (s_p^{fin})'(\frac{n}{L}) \sigma^2}}d\eta\\
&= \frac{1}{\sigma} \sqrt{1-i (s_p^{fin})'(\frac{n}{L}) \sigma^2} V_{\tilde f}^g(\frac{n}{L},s_p^{fin}(\frac{n}{L})).
\end{aligned}
\end{eqnarray}
Leading to the final estimation, recalling $(\frac{n}{L},s_p^{fin}(\frac{n}{L}))$ is not on the time-frequency grid: 
\begin{eqnarray}
\label{eq:retrieve-Mode2}
\begin{aligned}
f_p[n] \approx \frac{1}{\sigma} \sqrt{1-i (s_p^{fin})'(\frac{n}{L}) \sigma^2} V_{\tilde f}^g [n,k_0] e^{\frac{\pi \sigma^2(k_0-s_p^{fin}(\frac{n}{L}))^2}{1-i (s_p^{fin})'(\frac{n}{L}) \sigma^2}}
\end{aligned}
\end{eqnarray}
So the reconstruction formulae \eqref{eq:retrieve-Mode}  and \eqref{eq:retrieve-Mode2} are very close, since they are based on the same linear chirp approximation for the modes. 
Note also that a reconstruction formula similar to \eqref{eq:retrieve-Mode1} is used in \cite{li2020direct}, except that the STFT is replaced by the signal-separation operator which is 
a discrete version of the adaptive short-time Fourier transform studied for instance in \cite{li2020adaptive}. In that approach, the window parameter $\sigma$ is locally adapted to better separate the modes in the TF plane before reconstruction. A very interesting future development would certainly be to find a robust algorithm to adapt $\sigma$ 
locally so as to ease ridge detection with RRP-RD in very noisy situations, but this beyond the scope of the present article.      
  
The mode reconstruction technique based on \eqref{eq:retrieve-Mode}  can be adapted to S-RD and MB-RD, 
by replacing $s^{fin}_p(\frac{n}{L})$ by $\widehat{\omega}^{[2]}[n,\varphi_p[n]]$ (see \cite{laurent2020novel} for its definition),  
and $(s^{fin}_p)'(\frac{n}{L})$ by $\hat q_{\tilde f} [n,\varphi_p[n]]$ in \eqref{eq:approx_Vfp}. 
This technique is denoted by S-LCR-MR or by MB-LCR-MR when applied to S-RD or MB-RD, respectively. Note that  S-LCR-MR is 
exactly the technique programmed in \cite{laurent2020novel}.

\section{Numerical Applications}
\label{sec:res}
In this section, we first study the validity of the procedure for the determination of the number of modes introduced in Section \ref{sec:nummodes}, 
then compare  RRP-RD  with S-RD and MB-RD on simulated MCSs, in Section \ref{sec:res_ridge}, 
and evaluate the performance of the different mode retrieval techniques on these  signals, focusing on the role of ridge detection, in Section \ref{sec:res_rec}. 
We finally investigate the behavior of the reconstruction techniques based  on RRP-RD on a gravitational-wave signal, in Section \ref{sec:res_gravit}, 
and compare it with state-of-the-art techniques based on high-order synchrosqueezing transforms \cite{pham2017high}. 
Note that, as mentioned above, to compute STFT, in all cases we use a Gaussian window such that its standard deviation minimizes 
the R\'enyi entropy \cite{baraniuk2001measuring} of the TFR associated with STFT moduli.  
 We are aware of recent works on adaptive window determination \cite{li2020adaptive,li2019adaptive}, but though to choose the window adaptively may ease ridge determination, 
 such an approach is hard to carry out in noisy situations. In all the simulations we only consider negative input SNRs since at higher SNRs the ridge detection becomes less challenging. 
 \begin{figure*}[!htb]
\centering
\begin{minipage}{0.32\linewidth}
	\begin{tabular}{c}
	\includegraphics[width=\textwidth,height = 4 cm] {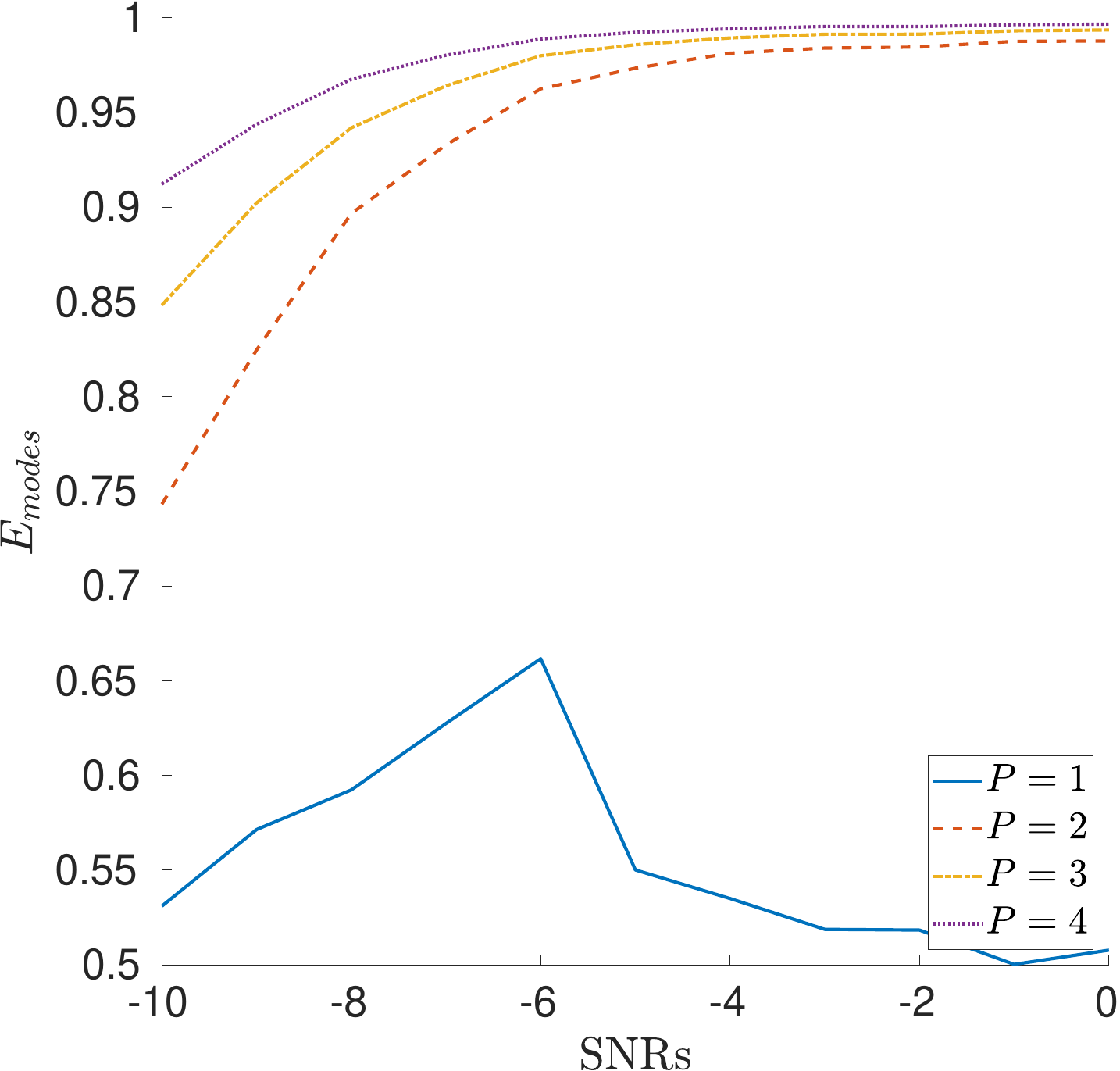}\\
	\hspace{0.6 cm} (a)
	\end{tabular}
\end{minipage}%
\begin{minipage}{0.32\linewidth}
	\begin{tabular}{c}
	\includegraphics[width=\textwidth,height = 4 cm] {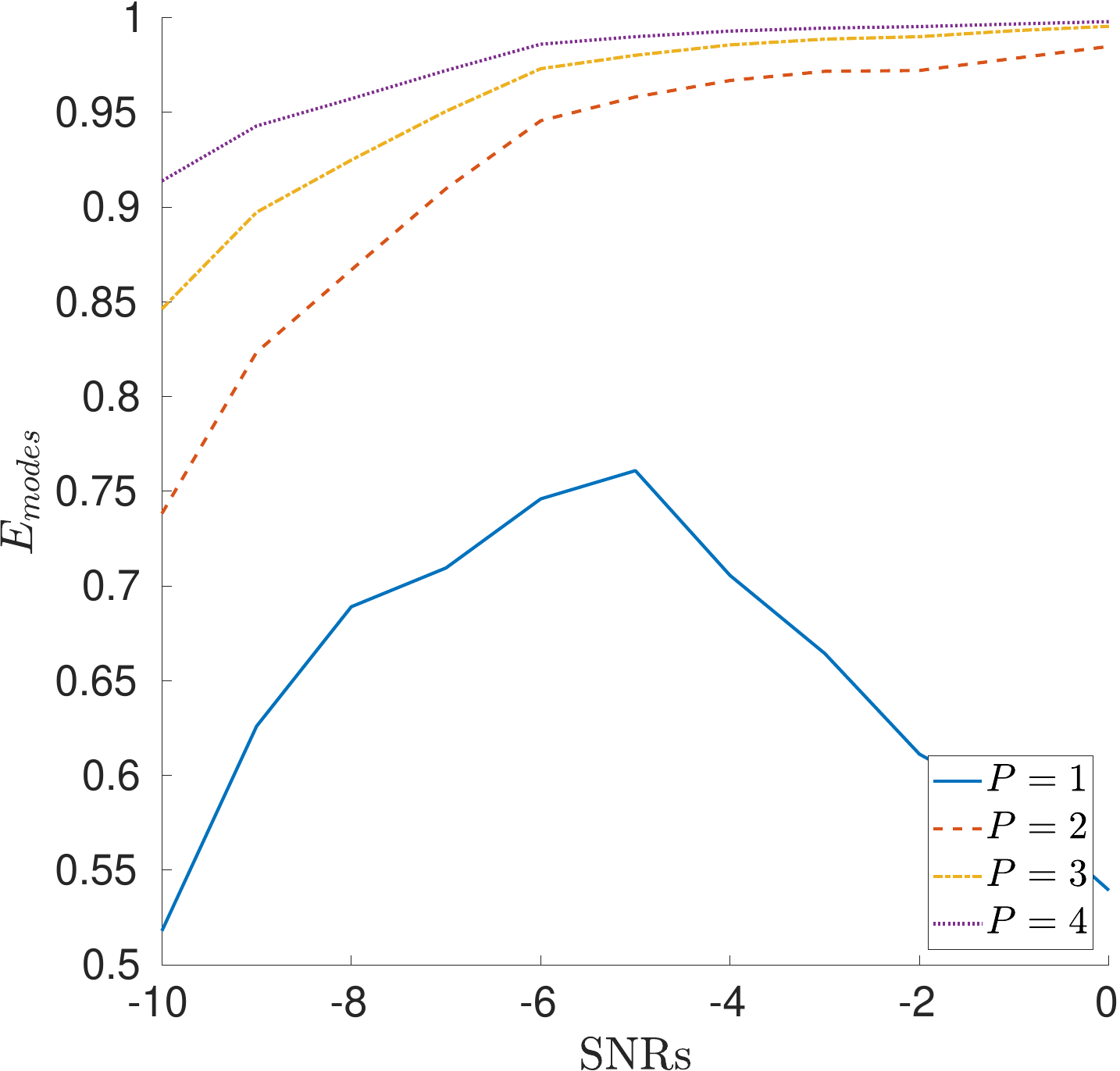}\\
    \hspace{0.6 cm} (b)
    \end{tabular}
\end{minipage}
\begin{minipage}{0.32\linewidth}
   \begin{tabular}{c}
	\includegraphics[width=\textwidth,height = 4 cm] {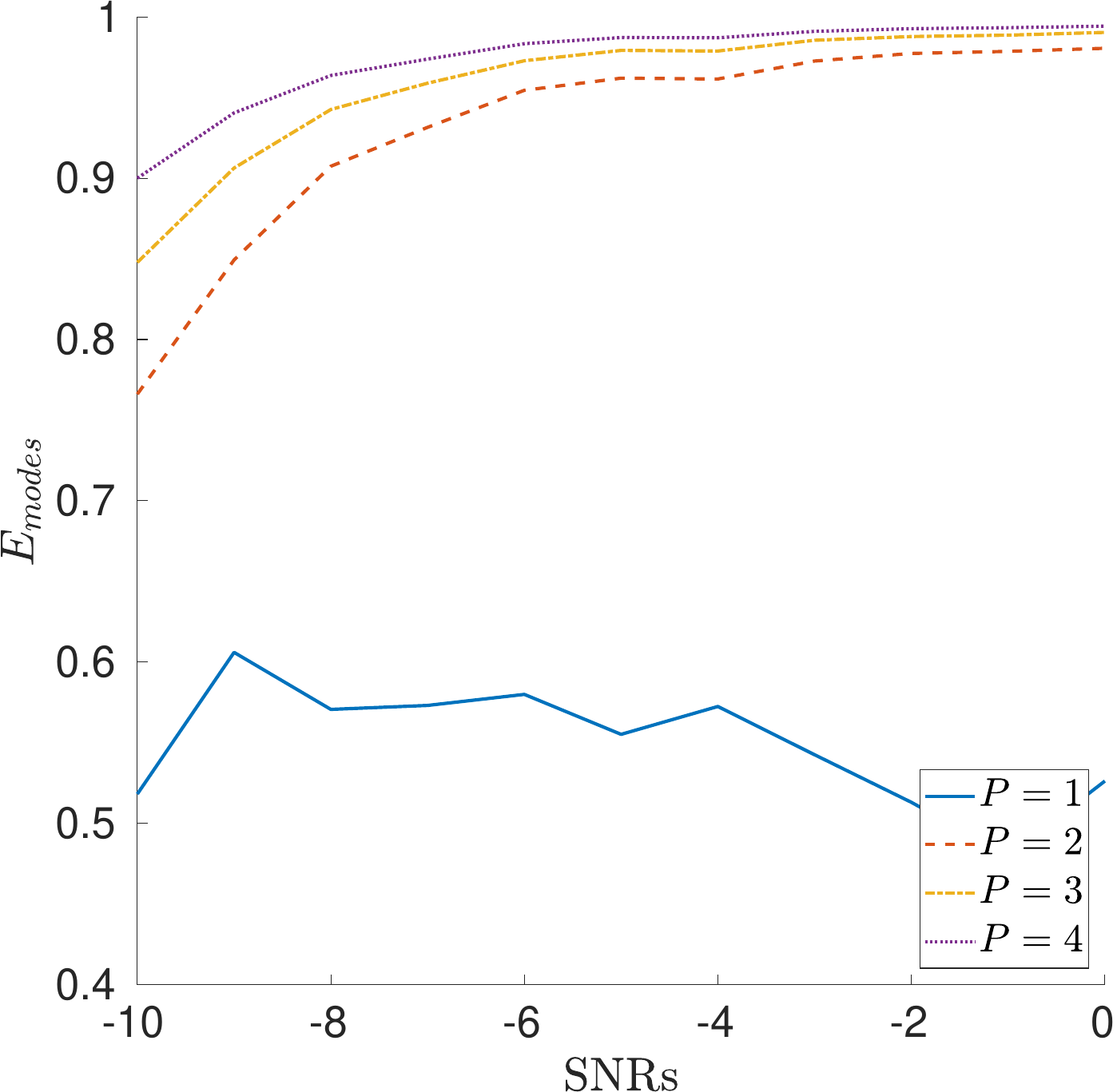}\\
    \hspace{0.6 cm} (c)
   \end{tabular}
\end{minipage}
\caption{(a): $E_{modes}(P)$ for the signal whose STFT modulus is displayed in Fig. \ref{Fig4} (a); 
(b): same as (a) but for the signal whose STFT modulus is displayed in Fig. \ref{Fig4} (b); 
(c):  same as (a) but for the signal whose STFT modulus is displayed in Fig. \ref{Fig4} (c). 
The results are averaged over 10 noise realizations.}
\label{Fig5}
\end{figure*}
    
\subsection{Evaluation of the Procedure to Determine the Number of Modes}
\label{sec:res_num_modes}
To investigate the procedure to determine the number of modes, we compute $E_{modes}$ introduced in Section \ref{sec:nummodes} 
for the signals whose STFTs are displayed in Fig. \ref{Fig4} (a), (b) and (c), when the noise level varies. The results displayed in  Fig. \ref{Fig5} 
show that $E_{modes}(1)$ is much lower than $1$, and that the difference between $E_{modes}$ evaluated at $P=2$ and $P=3$ gets smaller and smaller 
as the noise level decreases. Computing $\hat P$ as explained  in \eqref{def:hatP} leads to $\hat P = 2$ whatever the noise level.

\subsection{Comparison of RRP-RD, S-RD and MB-RD on Simulated Signals}
\label{sec:res_ridge}
Our goal in this section is to show that RRP-RD is more relevant in noisy situations than S-RD or MB-RD. 
For that purpose, we perform ridge detection for the signals whose STFTs are displayed on the first row of Fig. \ref{Fig4}, 
when the input SNR varies between -10 and 0 dB. 

\begin{figure*}[!htb]
\centering
\begin{minipage}{0.32\linewidth}
	\begin{tabular}{c}
	\includegraphics[width=\textwidth,height = 4 cm] {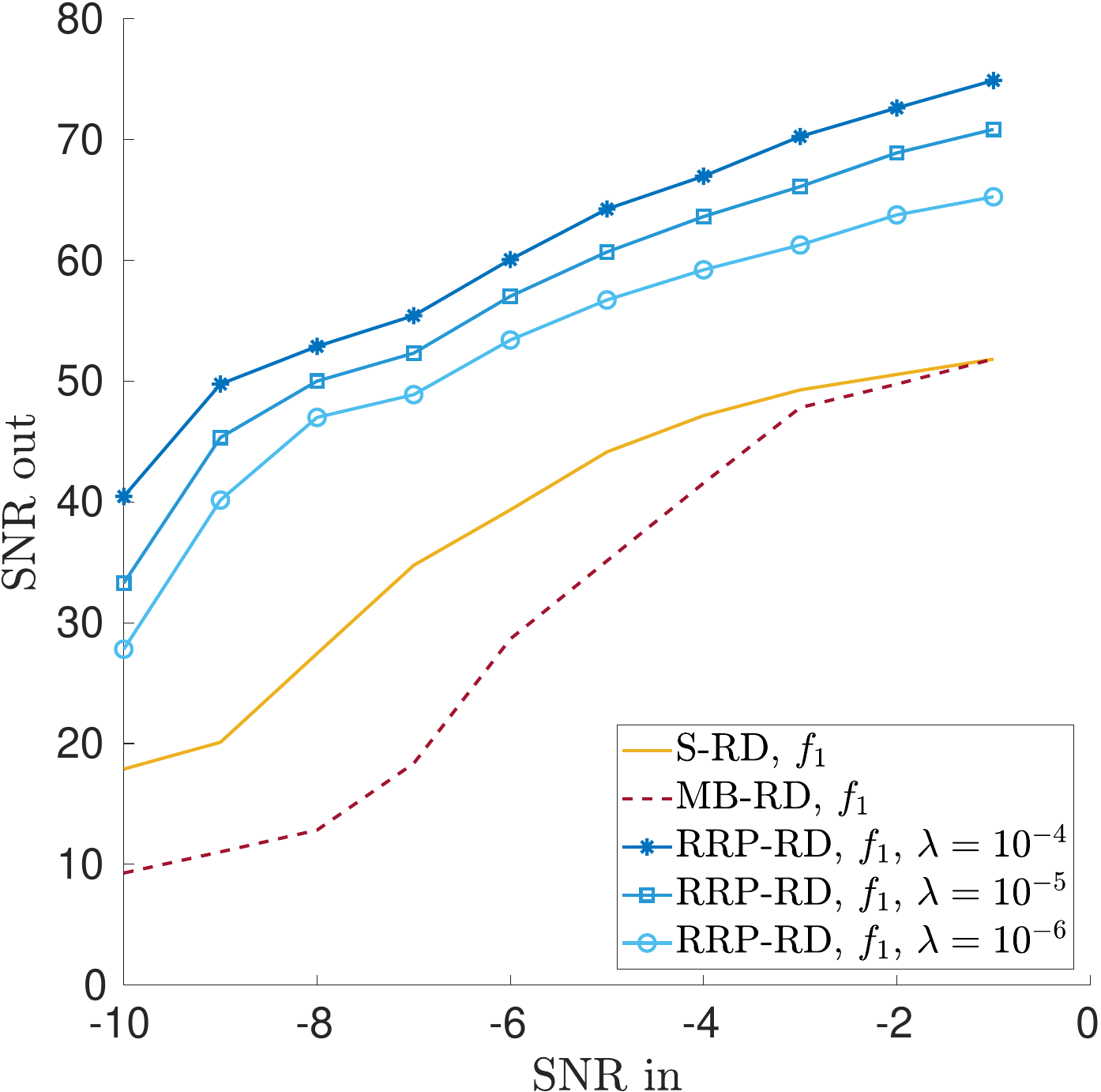}\\
	\hspace{0.6 cm} (a)
	\end{tabular}
\end{minipage}
\begin{minipage}{0.32\linewidth}
    \begin{tabular}{c}
	\includegraphics[width=\textwidth,height = 4 cm] {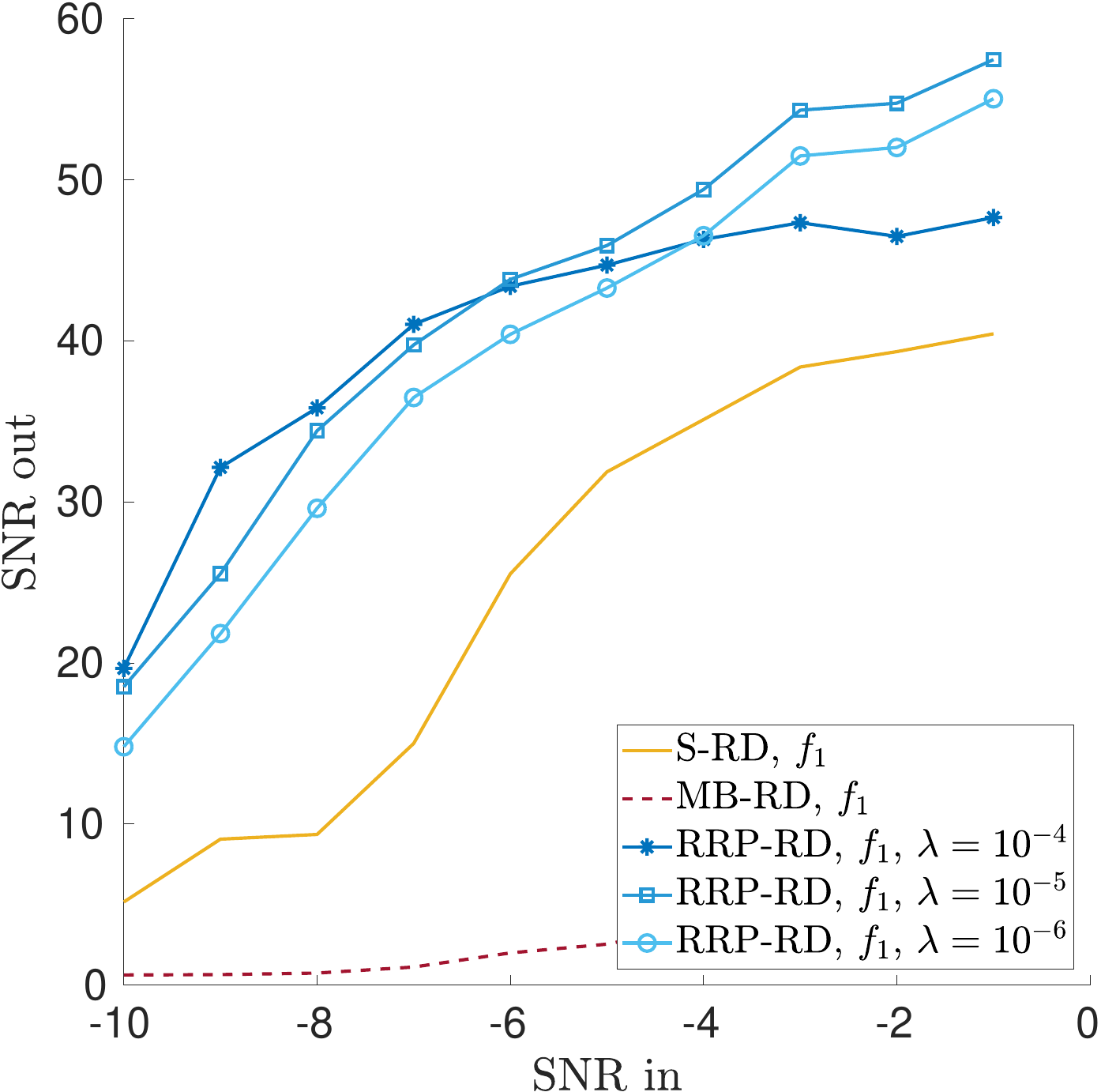}\\
    \hspace{0.6 cm} (b)
    \end{tabular}
\end{minipage}
\begin{minipage}{0.32\linewidth}
    \begin{tabular}{c}
	\includegraphics[width=\textwidth,height = 4 cm] {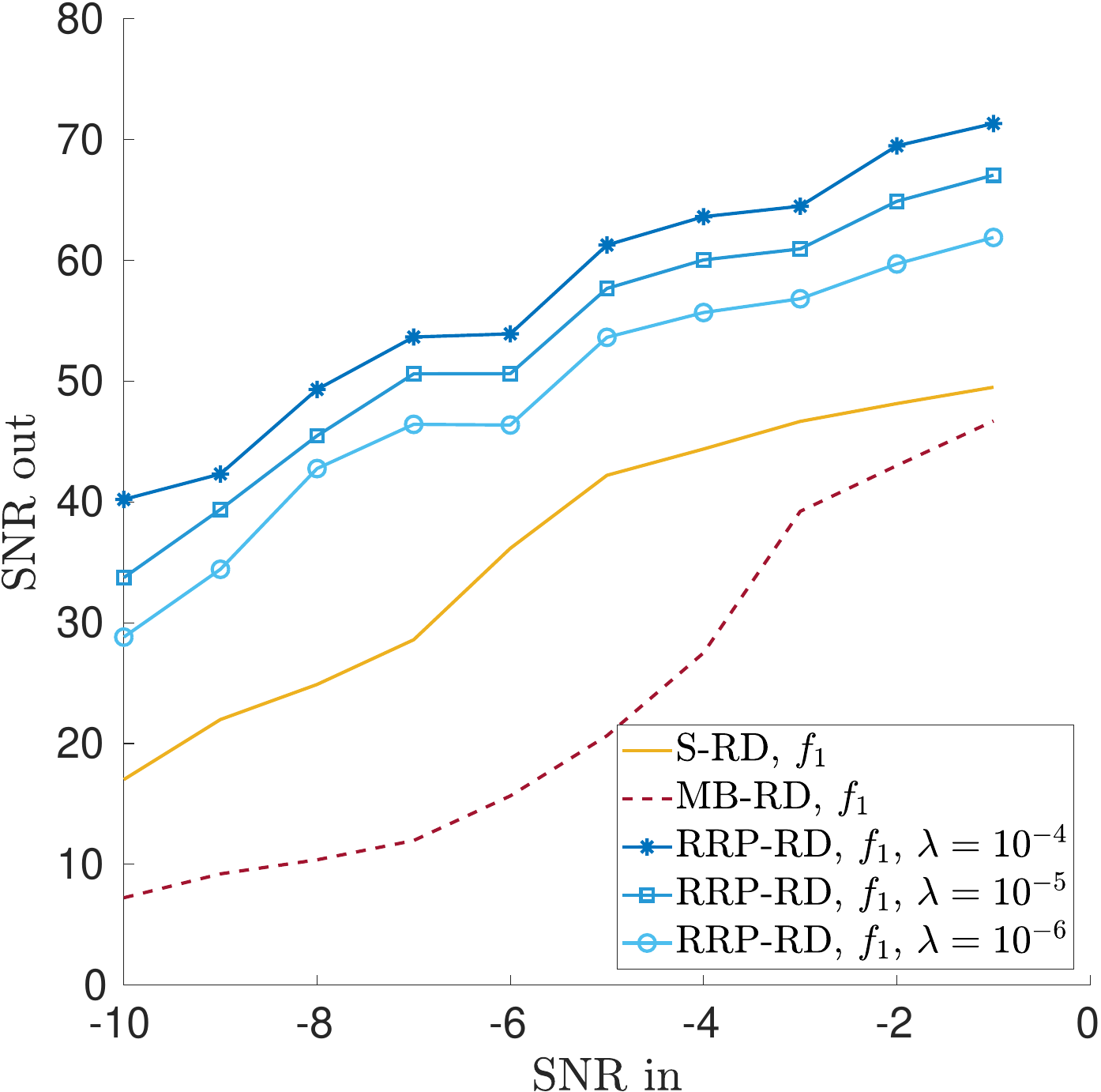}\\
    \hspace{0.6 cm} (c)
    \end{tabular}
\end{minipage}   
\centering
\begin{minipage}{0.32\linewidth}
	\begin{tabular}{c}
	\includegraphics[width=\textwidth,height = 4 cm] {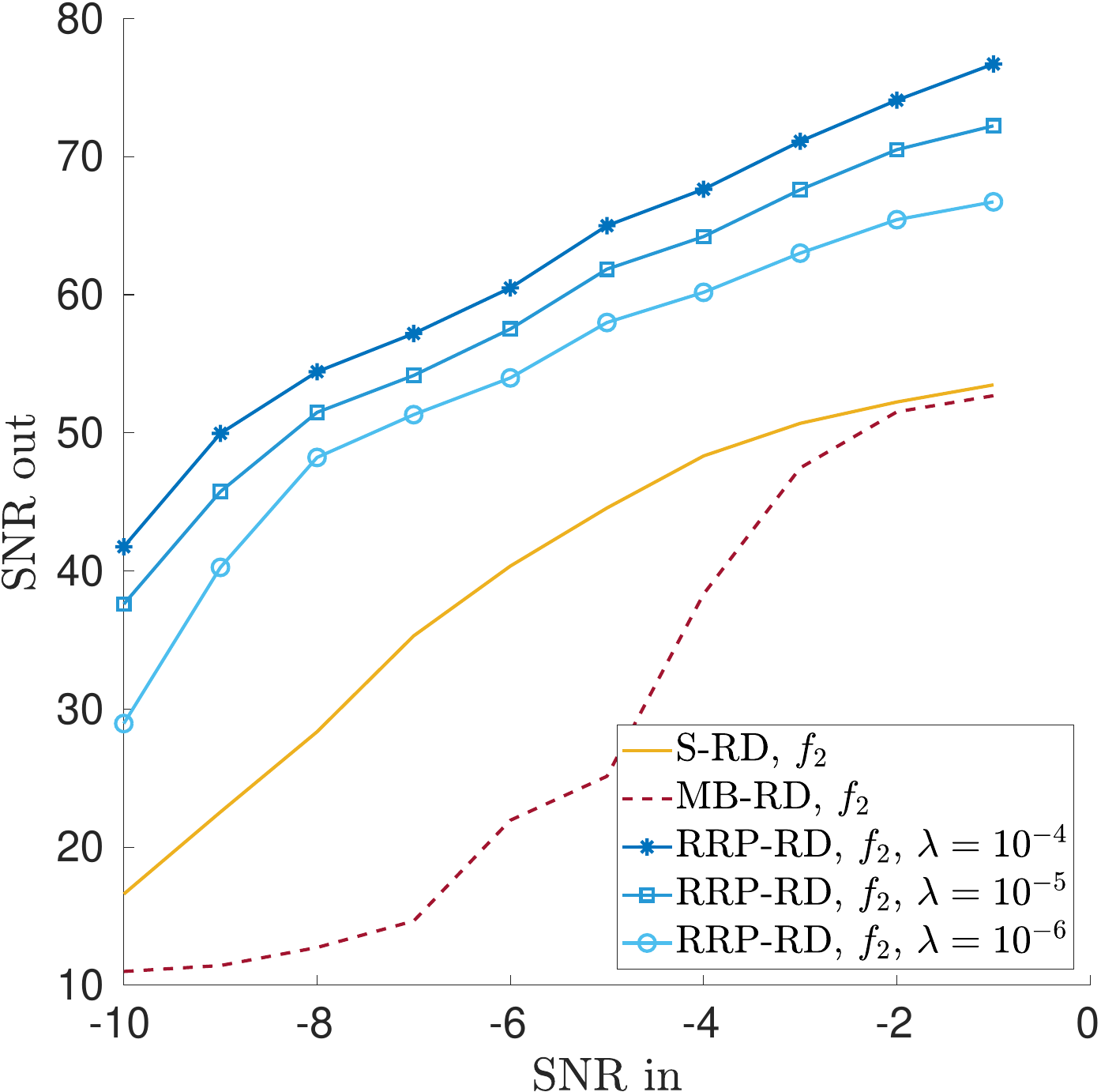}\\
	\hspace{0.6 cm} (d)
	\end{tabular}
\end{minipage}
\begin{minipage}{0.32\linewidth}
    \begin{tabular}{c}
	\includegraphics[width=\textwidth,height = 4 cm] {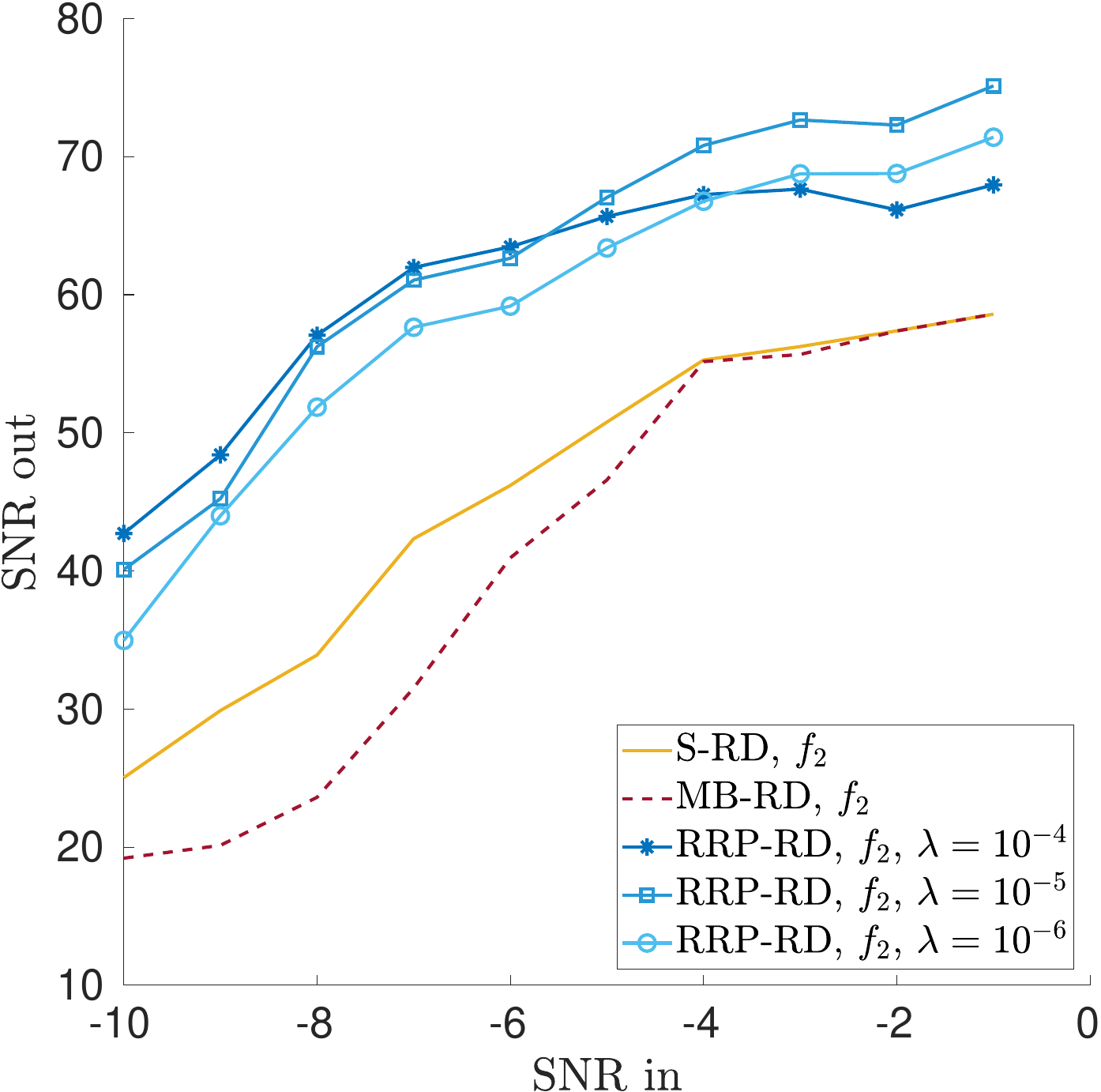}\\
    \hspace{0.6 cm} (e)
    \end{tabular}
\end{minipage}
\begin{minipage}{0.32\linewidth}
    \begin{tabular}{c}
	\includegraphics[width=\textwidth,height = 4 cm] {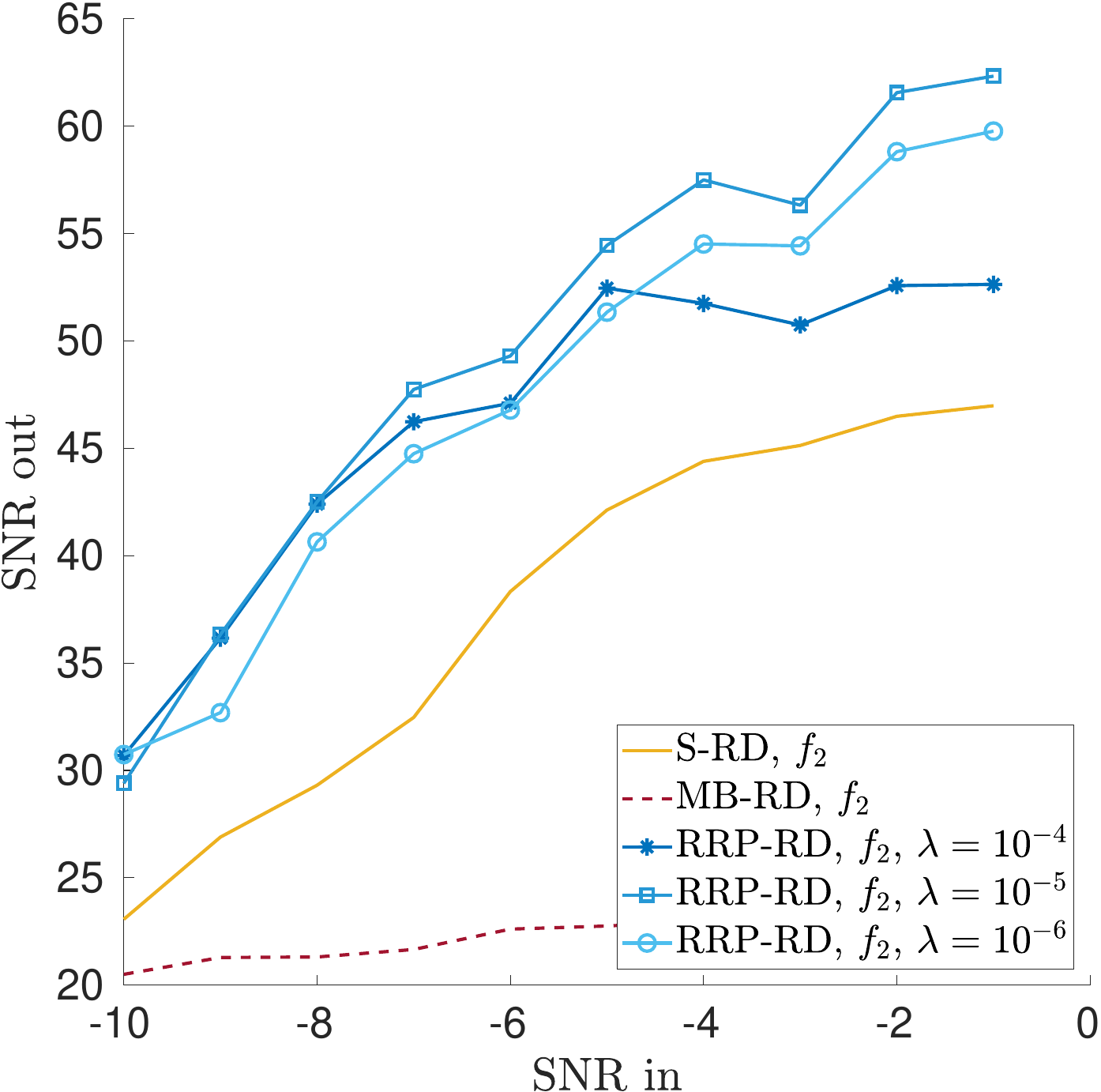}\\
    \hspace{0.6 cm} (f)
    \end{tabular}
\end{minipage}   

\caption{(a): Comparison between S-RD, MB-RD and RRP-RD, for 
the mode $f_1$ of the signal of Fig. \ref{Fig4} (a), 
computation of output SNR between IF $\phi_1'$ and estimated IF with respect to input SNR (the results are averaged over 40 noise realizations); (b): same as (a) but 
for the mode $f_1$ of  the signal of Fig. \ref{Fig4} (b); (c): same as (a) but for the mode $f_1$ of  the signal of Fig. \ref{Fig4} (c); 
(d): same as (a) but the mode $f_2$ of the signal of Fig. \ref{Fig4} (a); 
(e): same as (b) but the mode $f_2$ of the signal of Fig. \ref{Fig4} (b);
(f): same as (c) but the mode $f_2$ of the signal of Fig. \ref{Fig4} (c).} 
\label{Fig6}
\end{figure*}

For the two linear chirps signal of Fig. \ref{Fig4} (a), the ridge detection results are depicted in Fig. \ref{Fig6} (a)  and (d) for mode $f_1$ and $f_2$, respectively. 
These simulations first tell us that RRP-RD performs much better than S-RD and MB-RD, the results being very similar for the two modes. 
It is worth remarking that to consider a higher smoothing parameter $\lambda$ in RRP-RD enables better ridge detection for that type of signals. 
Then, comparing the results for  S-RD and MB-RD, we notice that the former behaves better than the latter, especially at high noise level.
Indeed, to use only  the modulation operator $\hat q_{\tilde f}$ for ridge detection leads to inaccuracies as soon the ridge is split, therefore MB-RD 
fails to follow the different ridge portions corresponding to a mode (in these simulations, $C$ is set to $2$). 
On the contrary, since S-RD uses the fixed modulation parameter $B_f$ (here set to $10$), it is able to better follow disconnected ridge portions.
At higher input SNRs, S-RD and MB-RD lead to very similar results, since, in these cases, the LMMFs corresponding to the two largest STFT 
modulus maxima are, for most time indices, close to the true IF locations of the modes. 

The ridge detection results for the signal of Fig. \ref{Fig4} (b), displayed in Fig. \ref{Fig6} (b) and (e) for modes $f_1$ and $f_2$ respectively, 
tell us that the behavior of RRP-RD on mode $f_2$ is similar to that on a linear chirp: RRP-RD outperforms the two other tested techniques. 
The only difference is that a larger smoothing parameter in RRP-RD leads to better results but only when the noise level is very high. 
As for mode $f_1$, which is much more modulated that $f_2$, RRP-RD is still much better than the other two techniques, 
and we remark that MB-RD does not achieve ridge detection when high noise is combined with strong frequency modulation. 
\begin{figure*}[!htb]
\centering
\begin{minipage}{0.32\linewidth}
	\begin{tabular}{c}
	\includegraphics[width=\textwidth,height = 8 cm] {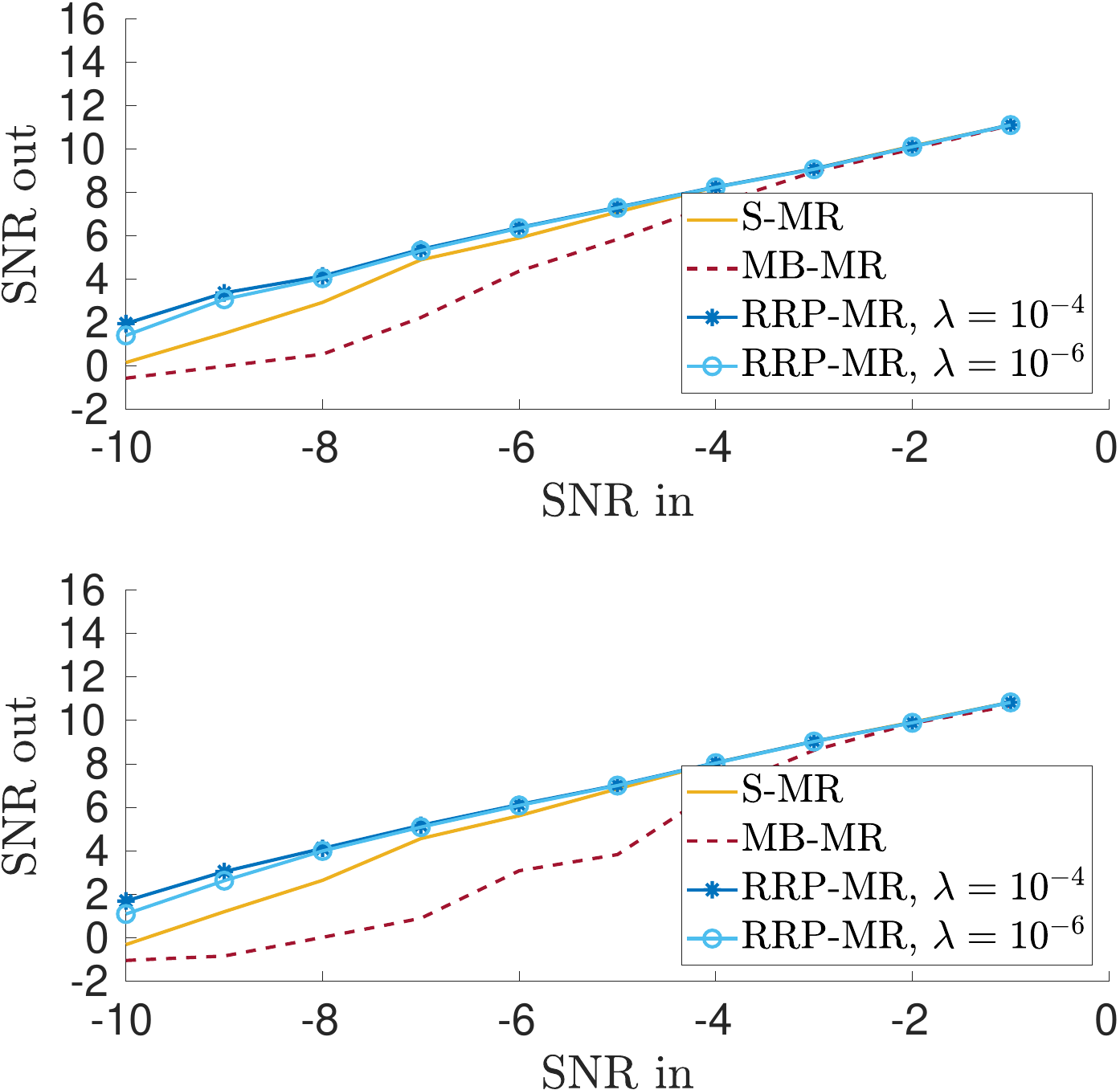}\\
	\hspace{0.6 cm} (a)
	\end{tabular}
\end{minipage}
\begin{minipage}{0.32\linewidth}
    \begin{tabular}{c}
	\includegraphics[width=\textwidth,height = 8 cm] {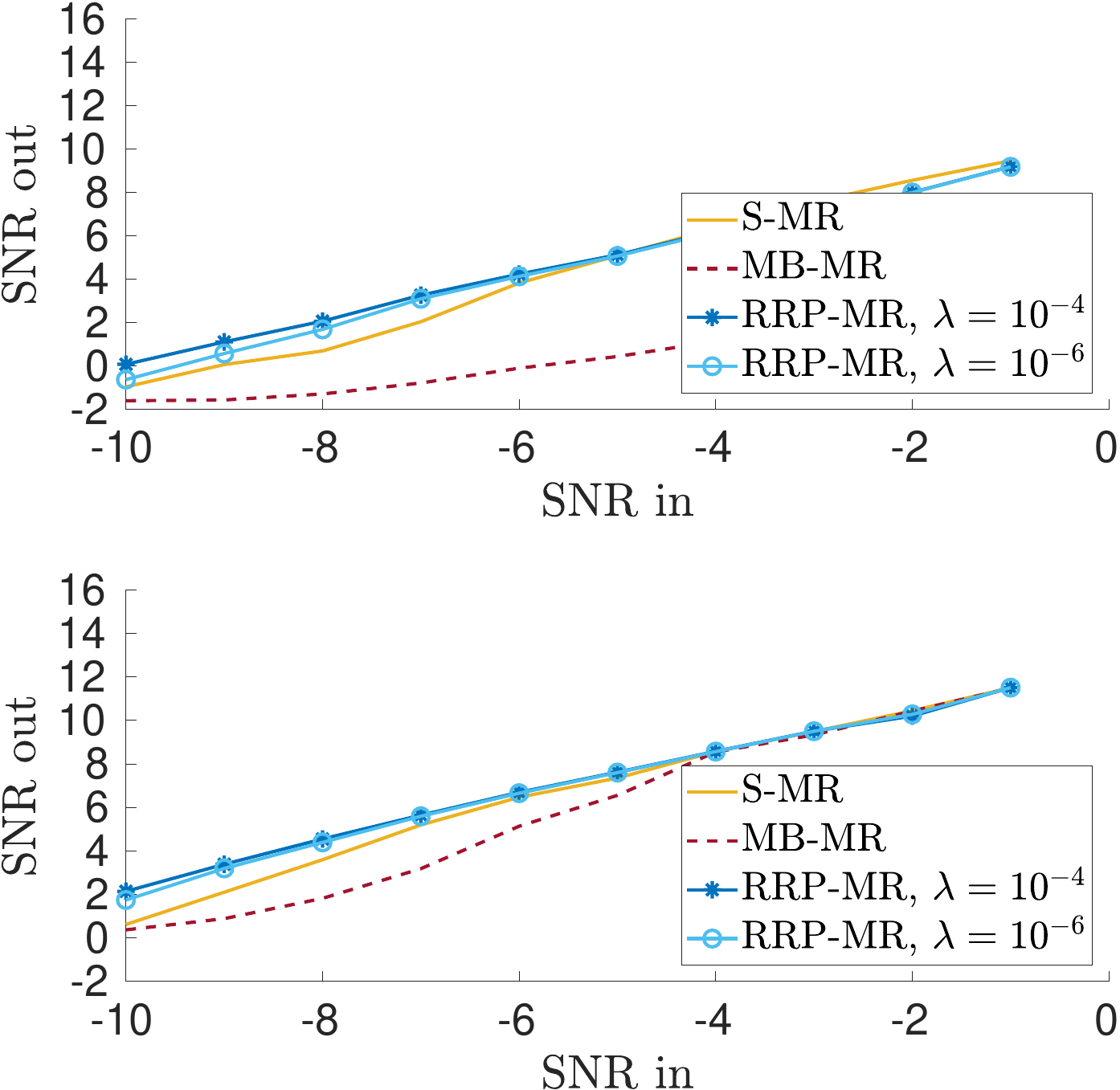}\\
    \hspace{0.6 cm} (b)
    \end{tabular}
\end{minipage}
\begin{minipage}{0.32\linewidth}
    \begin{tabular}{c}
	\includegraphics[width=\textwidth,height = 8 cm] {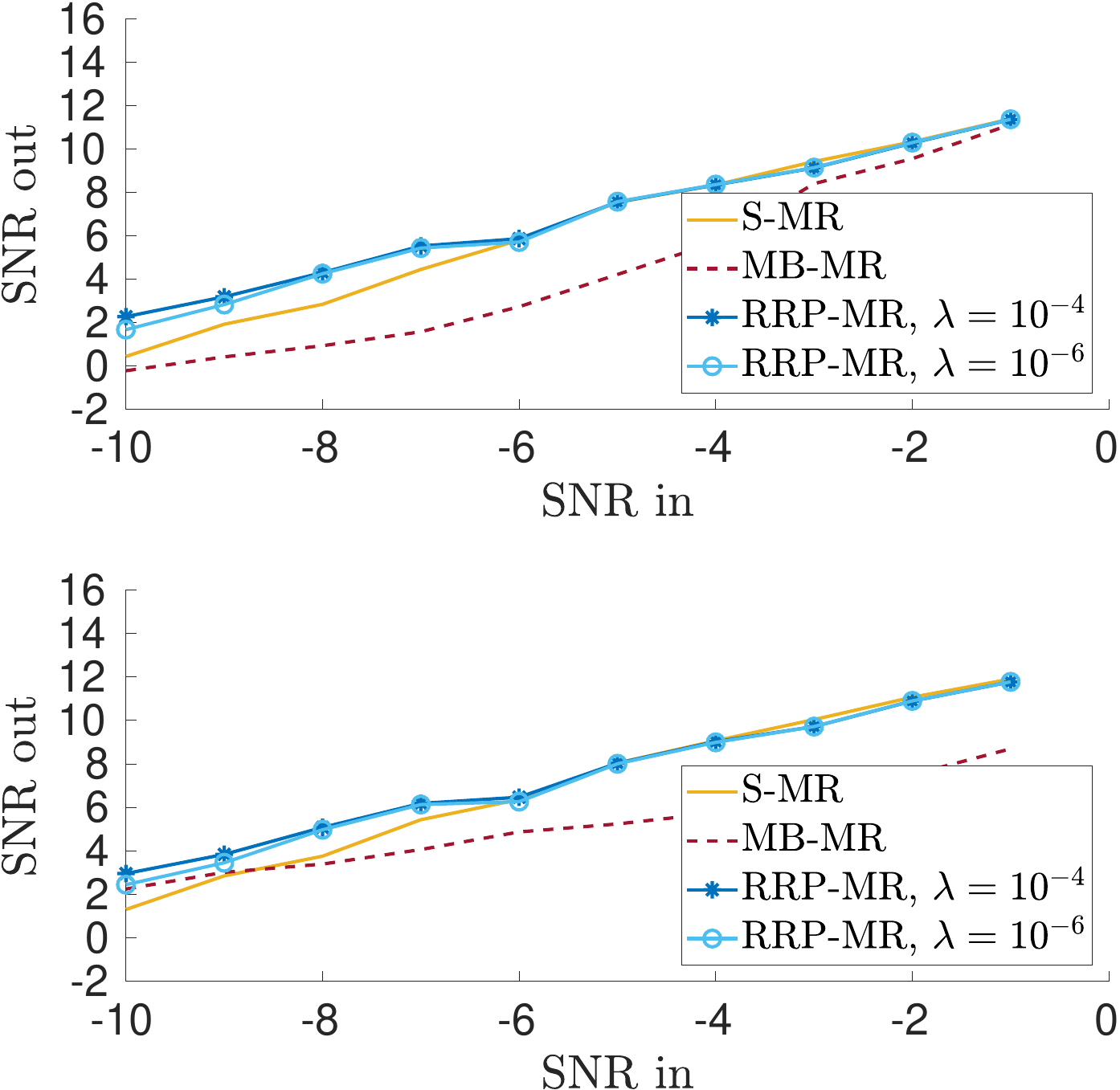}\\
    \hspace{0.6 cm} (c)
    \end{tabular}
\end{minipage}
\caption{(a): For each mode $p = 1, 2$, output SNR between mode 
$f_p$ of signal of Fig. \ref{Fig4} (a) and reconstructed mode for 
each methods, namely S-MR, MB-MR, RRP-MR (top: mode $f_1$, bottom: mode $f_2$).  
The results are averaged over 40 noise realizations;
(b): same but with signal of Fig. \ref{Fig4} (b);
(c): same but with signal of Fig. \ref{Fig4} (c);}
\label{Fig7}
\end{figure*}
\begin{figure*}[!htb]
\centering
\begin{minipage}{0.32\linewidth}
	\begin{tabular}{c}
	\includegraphics[width=\textwidth,height = 4 cm] {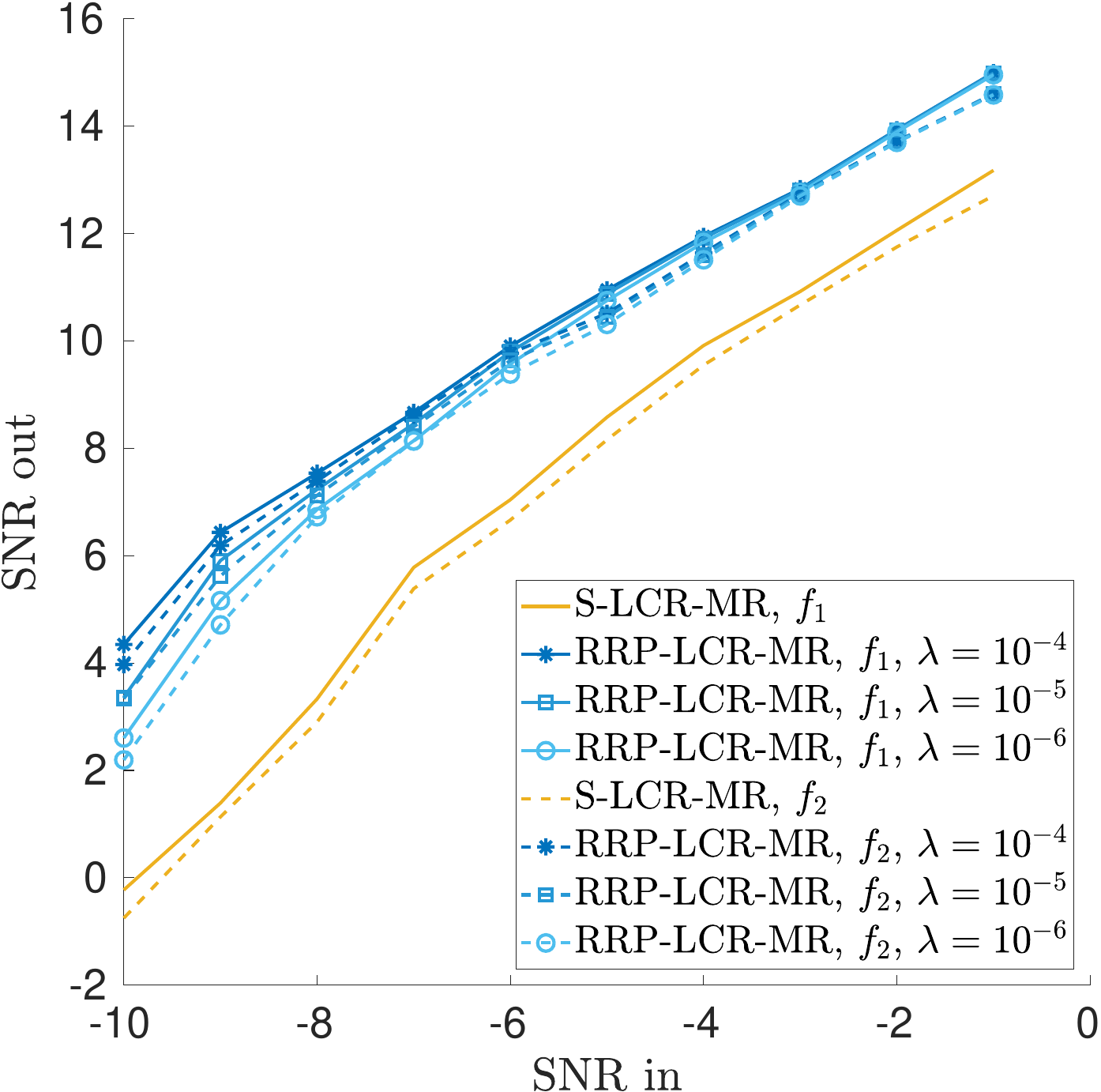}\\
	\hspace{0.6 cm} (a)
	\end{tabular}
\end{minipage}
\begin{minipage}{0.32\linewidth}
    \begin{tabular}{c}
	\includegraphics[width=\textwidth,height = 4 cm] {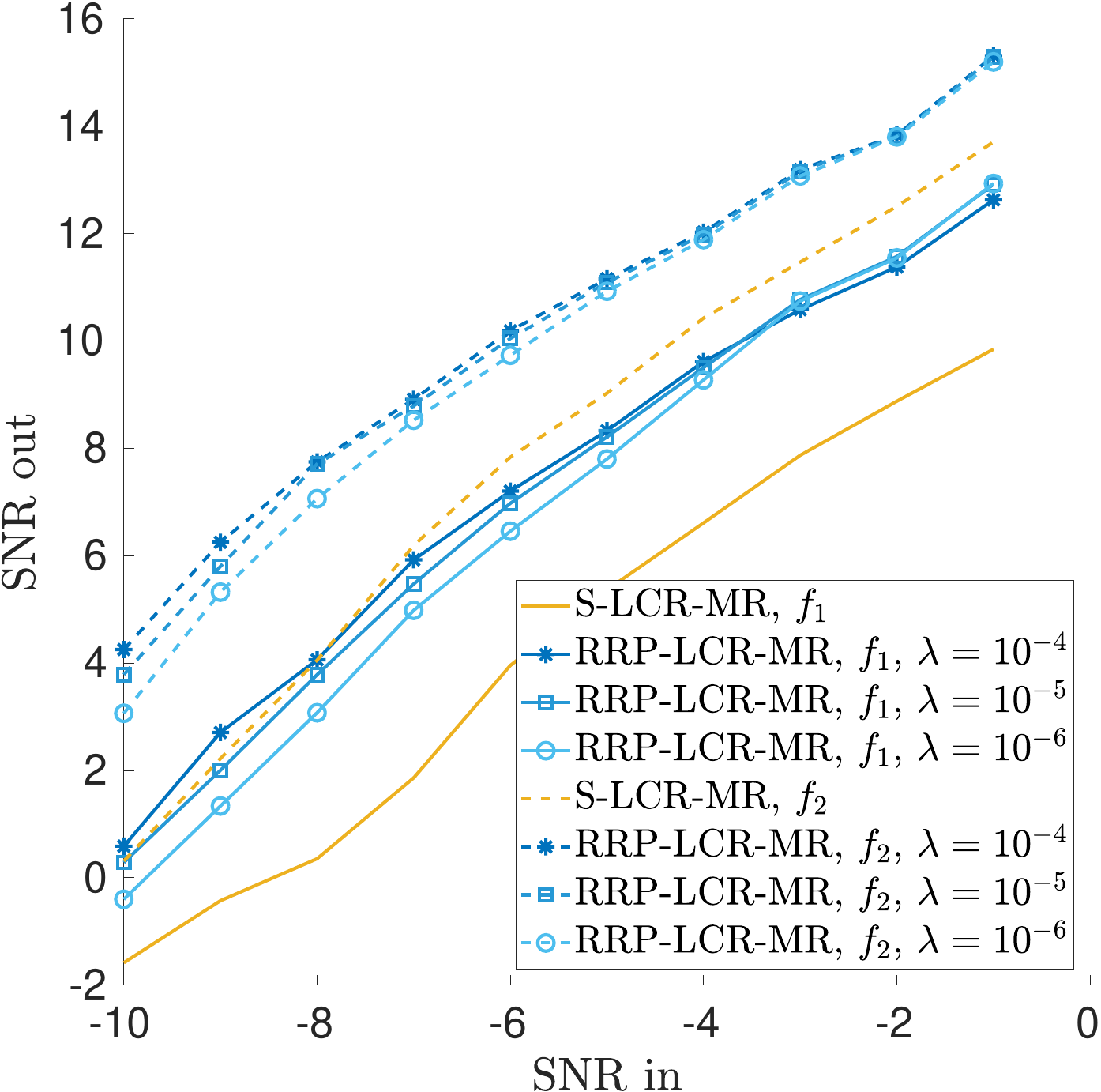}\\
    \hspace{0.6 cm} (b)
    \end{tabular}
\end{minipage}
\begin{minipage}{0.32\linewidth}
    \begin{tabular}{c}
	\includegraphics[width=\textwidth,height = 4 cm] {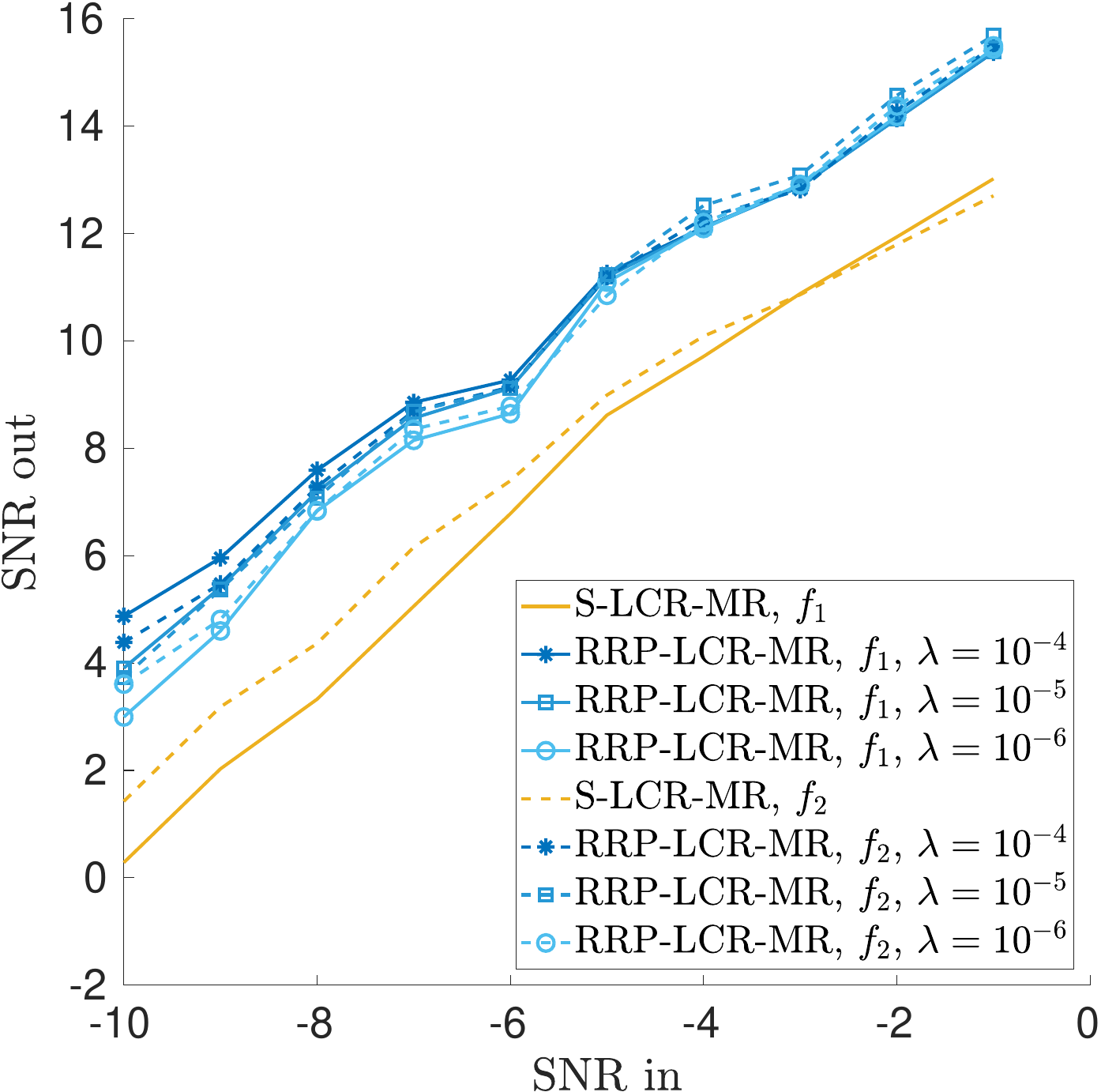}\\
    \hspace{0.6 cm} (c)
    \end{tabular}
\end{minipage}
\caption{(a): For each mode $p = 1, 2$, output SNR between mode 
$f_p$ of signal of Fig. \ref{Fig4} (a) and reconstructed mode using either 
S-LCR-MR or RRP-LCR-MR (the results 
are averaged over 40 noise realizations);
(b): same but with signal of Fig. \ref{Fig4} (b);
(c): same but with signal of Fig. \ref{Fig4} (c);}
\label{Fig8}
\end{figure*}

Finally, the ridge detection results for the signal of Fig. \ref{Fig4} (c) are displayed in Fig. \ref{Fig6} (c) and (f) for modes $f_1$ and $f_2$ respectively, and 
we only comment on ridge detection for $f_2$ which has an exponential phase:  we again notice that RRP-RD behaves much better than the other two tested methods, that the smoothing parameter should be chosen all the larger the higher the noise level is, and that MB-RD is not competitive in that case, for the same reason as before.

\subsection{Comparison of Mode Retrieval Techniques}
\label{sec:res_rec}
In this section, we investigate the quality of mode retrieval techniques S-MR, MB-MR, RRP-MR, on the one hand, 
and, on the other hand, S-LCR-MR, MB-LCR-MR and RRP-LCR-MR, still for the signals displayed on the first row of Fig. \ref{Fig4}.

Looking at the results of Fig. \ref{Fig7} (a) related to the signal 
of Fig. \ref{Fig4} (a), it transpires that while RRP-RD is much better than S-RD and MB-RD  this improvement is 
not as significant in the associated mode reconstruction techniques.  We can however remark that RRP-MR always behaves better than the other tested 
methods, that the smoothing parameter $\lambda$ used in RRP-RD seems to have very little influence on mode reconstruction, and 
that MB-MR behaves always worse since its performance are hampered by inaccurate ridge detection.
So, good ridge detection does not warranty good mode reconstruction, and such a conclusion remains valid when one applies S-MR, MB-MR and RRP-MR 
to the signals of Fig. \ref{Fig4} (b) and (c). 
This means that even if RRP-RD finds the right TF location for the modes, the coefficients in the vicinity of the ridges are too damaged by noise to enable an accurate mode retrieval by summing the coefficients in the TF plane.

\begin{figure}[!htb]
\centering
\begin{tabular}{c c}
\includegraphics[width = 6 cm,height = 5 cm] {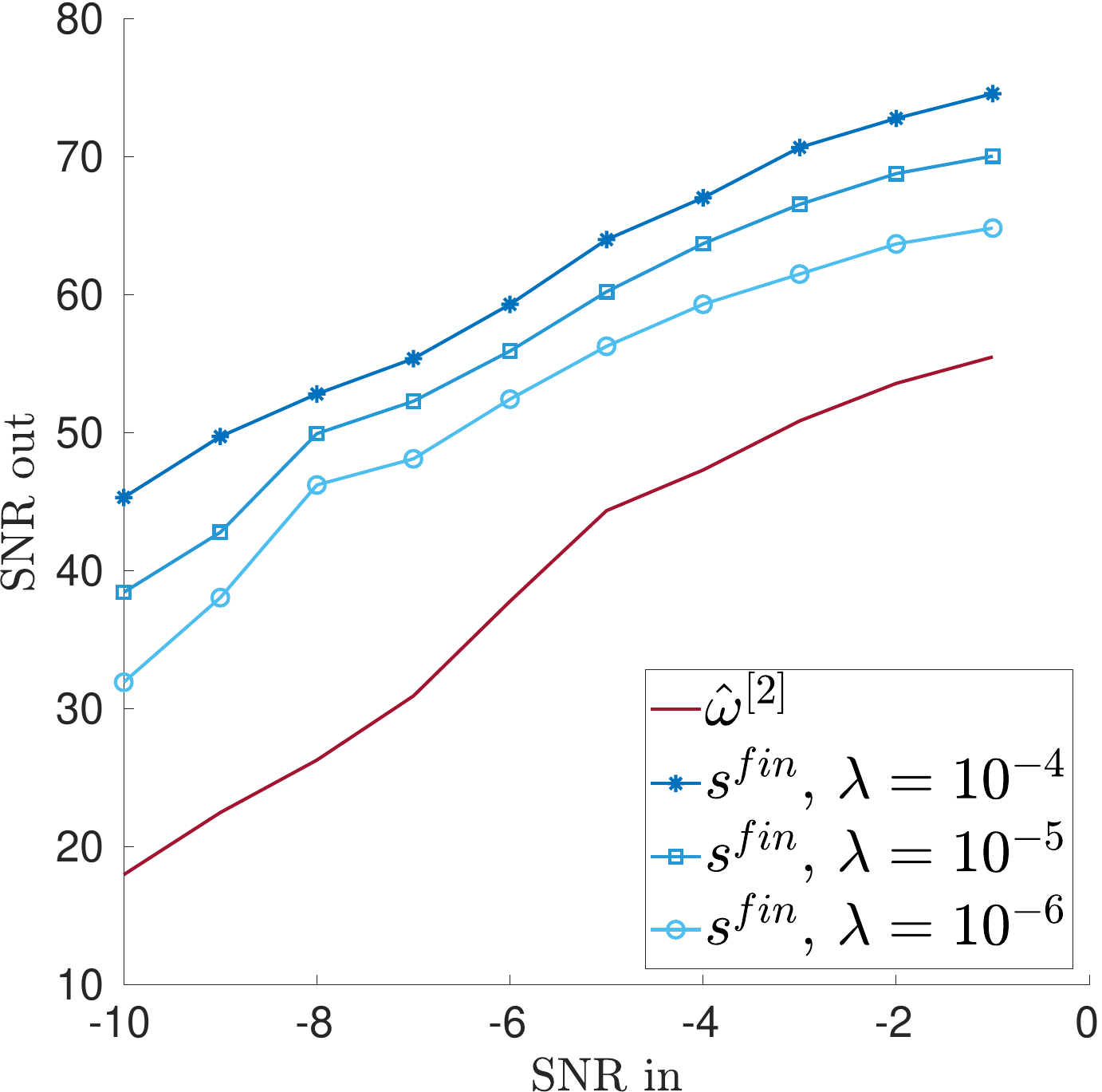}&
\includegraphics[width = 6 cm,height = 5 cm] {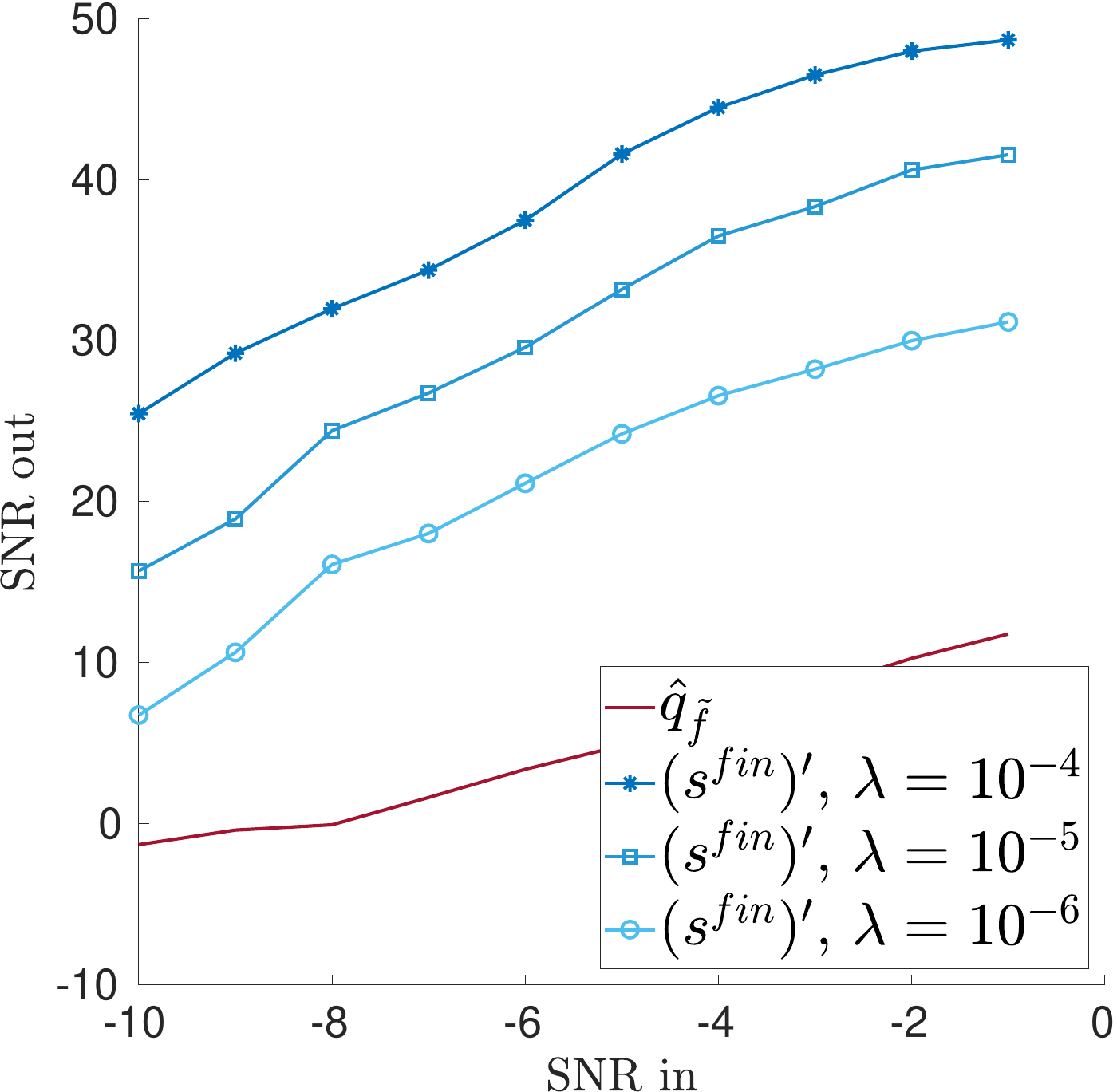}\\
 (a) & (b)
 \end{tabular}
\caption{(a): computation of the output SNR associated with the estimation of $\phi'$ with $s^{fin}$ (computed for different values of $\lambda$) or with $\widehat{\omega}^{[2]}$ for the first mode of Fig. \ref{Fig4} (a); (b): computation of the SNR associated with the estimation of $\phi''$ with $(s^{fin})'$ (computed for different values of $\lambda$) or with $\widehat{q}_{\tilde f}$ for the first mode of Fig. \ref{Fig4} (a).}
\label{Fig8bis}
\end{figure}

The results of Fig. \ref{Fig8} compared with those of Fig. \ref{Fig7} first show the superiority of RRP-LCR-MR over RRP-MR: to consider a linear chirp approximation in
the vicinity of the detected ridges is more relevant than to sum the coefficients in the TF plane.  Then, looking at Fig. \ref{Fig8} only, we notice 
that  RRP-LCR-MR behaves better than the original LCR technique introduced in \cite{laurent2020novel} (S-LCR-MR in the present paper). The reason for 
such an improvement is that  $s_p^{fin}$ and  $(s_p^{fin})'$ are better estimators of  $\phi_p'$ and $\phi_p''$ than  
$\widehat{\omega}^{[2]}$ and $\hat q_{\tilde f}$ evaluated on the ridges given by S-RD. To confirm this, we display  in Fig. \ref{Fig8bis} the SNRs associated with the estimation 
of $\phi'$ and $\phi''$ by $s^{fin}$ and  $(s^{fin})'$ or by $\widehat{\omega}^{[2]}$ and $\hat q_{\tilde f}$, for the first mode of Fig. \ref{Fig4} (a).  
Going back Fig. \ref{Fig8} we do not display the mode reconstruction results associated with MB-LCR-MR since these are significantly 
worse than those presented here. Finally we shall mention that the quality of mode reconstruction with RRP-LCR-MR depends only very slightly  
on the value of the smoothing parameter, and that, with this technique, the quality of mode reconstruction is very similar for most types of modes. 

\subsection{Application to Gravitational-Wave Signals}
\label{sec:res_gravit}
In this section, we investigate the applicability of RRP-RD and RRP-LCR-MR  to a transient gravitational-wave signal, 
generated by the coalescence of two stellar-mass black holes. 
This event, called \textbf{GW150914}, was detected by the LIGO detector Hanford, Washington and closely matches the waveform Albert Einstein predicted almost 100 years ago in his general relativity theory for the inspiral, the merger of a pair of black holes and the ringdown
of the resulting single black hole \cite{Abbott2016}.  
The observed signal has a length of 3441 samples in $T = 0.21$ seconds.
\begin{figure*}[!htb]
\begin{minipage}{0.33\linewidth}
	\begin{tabular}{c}
	\includegraphics[width=\textwidth,height = 4 cm] {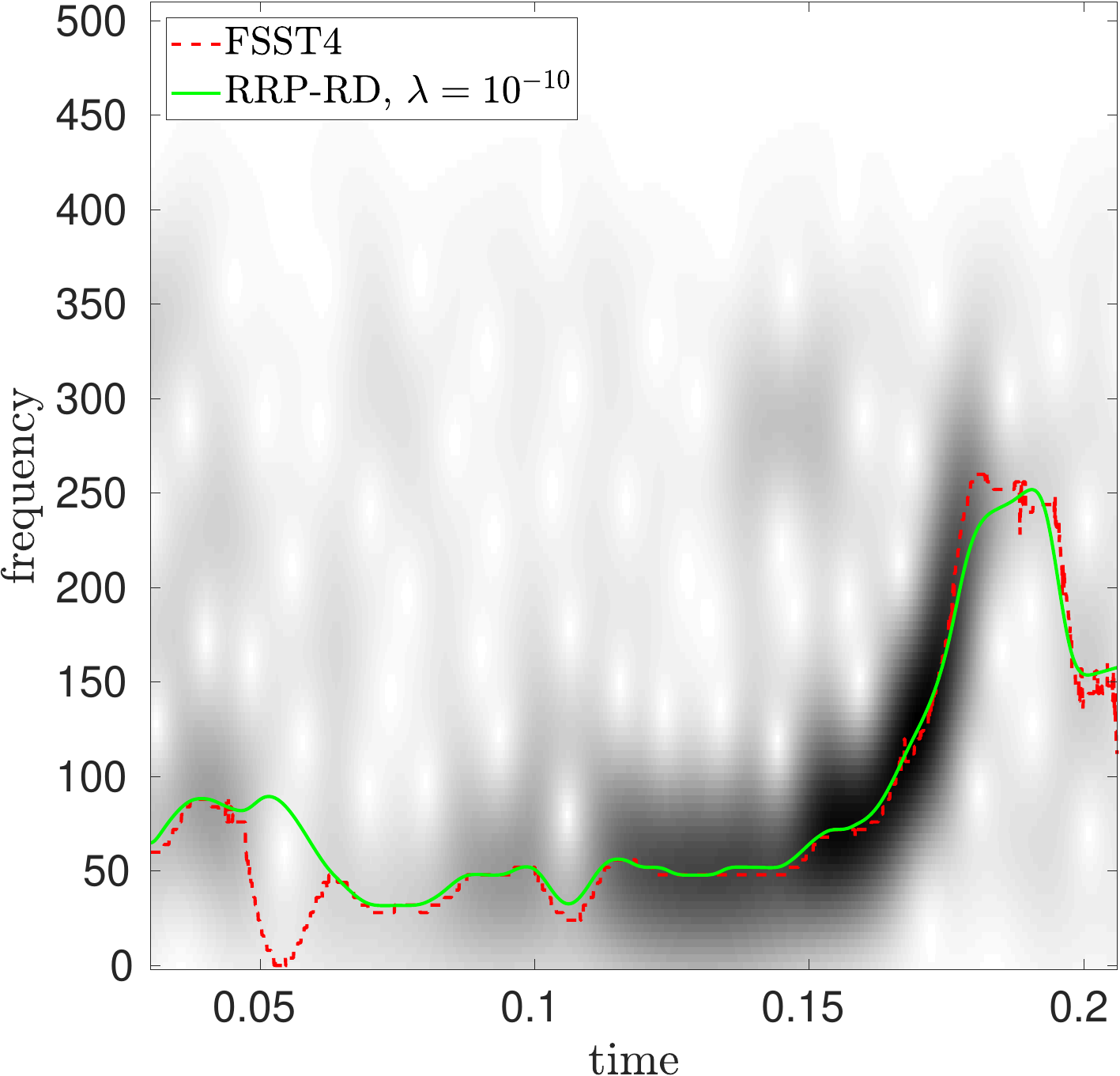}\\
	\hspace{0.6 cm} (a)
	\end{tabular}
\end{minipage}
\begin{minipage}{0.33\linewidth}
	\begin{tabular}{c}
	\includegraphics[width=\textwidth,height = 4 cm] {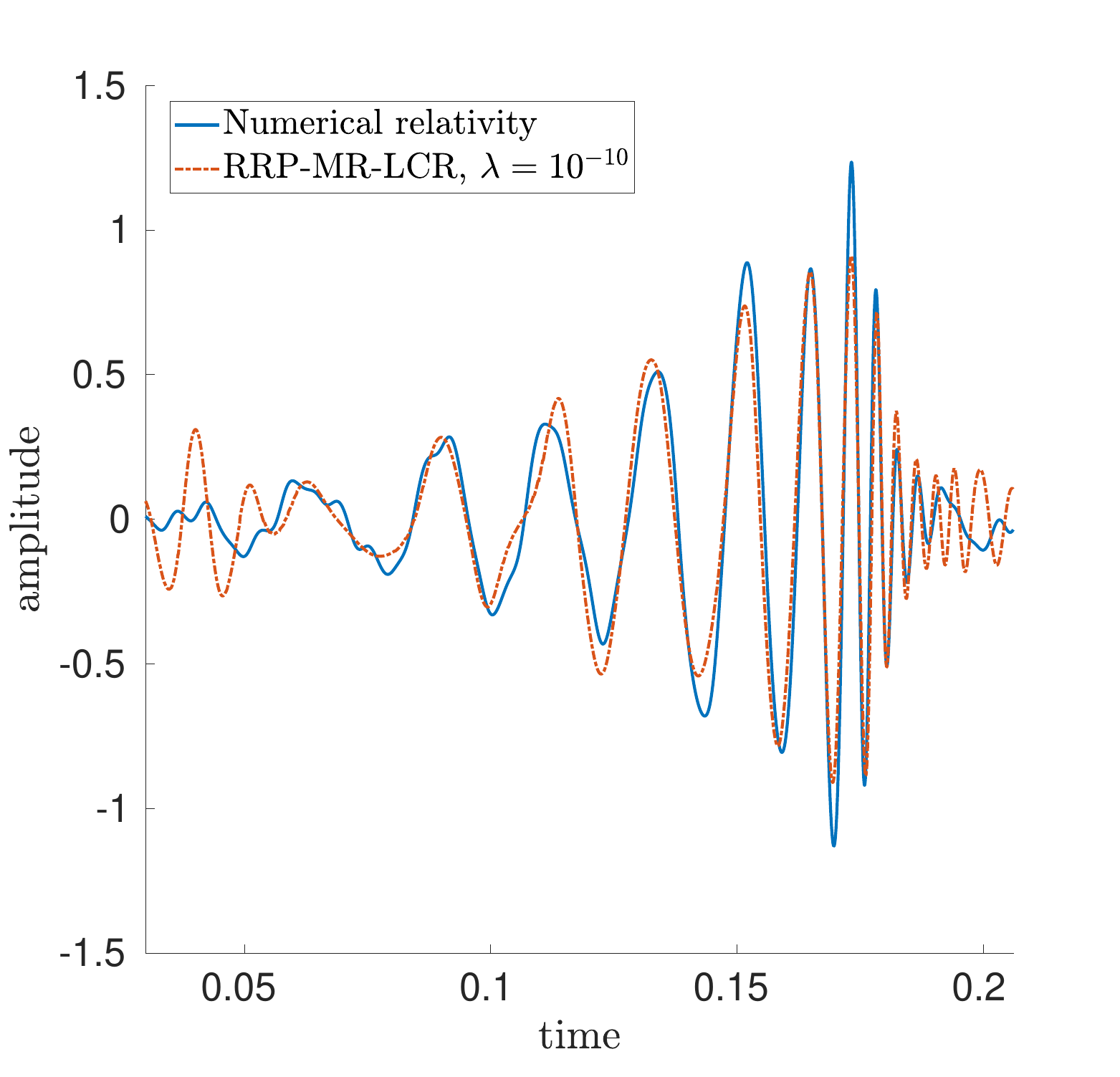}\\
	\hspace{0.6 cm} (b)
	\end{tabular}
\end{minipage}
\begin{minipage}{0.33\linewidth}
	\begin{tabular}{c}
\includegraphics[width=\textwidth,height = 4 cm] {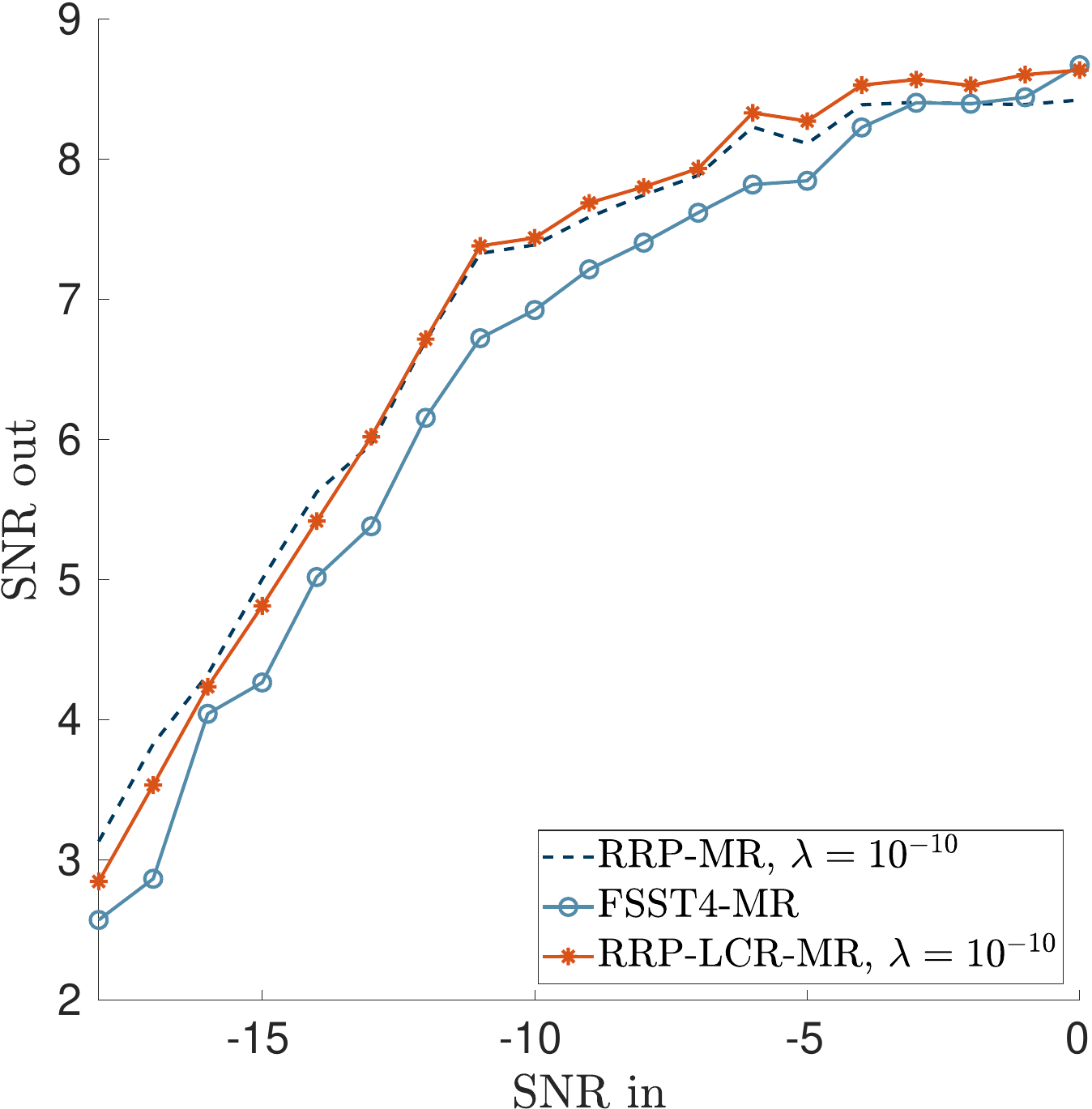}\\
	\hspace{0.6 cm} (c)
	\end{tabular}
\end{minipage}
\caption{(a): STFT modulus ($\sigma= 0.0105$) of the Hanford signal along with the ridge obtained with RRP-RD and FSST4; (b): 
iIlustration of  signal reconstruction based on RRP-LCR-MR and the numerical relativity;
(c): SNR corresponding to the reconstruction of the signal using either RRP-LCR-MR, RRP-MR or FSST4-MR (the ground truth is assumed to be the one produced by numerical relativity). The results are average over 40 noise realizations.}
\label{Fig9}
\end{figure*}

We first display in Fig. \ref{Fig9} (a), the modulus of the STFT of such a signal, along with the spline obtained 
using RRP-RD. For the sake of comparison, we also display the ridge obtained by considering 
the fourth-order synchrosqueezing transform, an efficient reassignment technique introduced in \cite{pham2017high}. 
Such a ridge is denoted by FSST4 in that figure. We notice that RRP-RD and FFST4 leads to very similar results, and 
that  both techniques  enables the detection of the ringdown.  

We then investigate the quality of signal reconstruction by comparing  
it with the one given by the numerical relativity \cite{abbott2016gw151226}, when one uses either  RRP-MR, RRP-LCR-MR or the mode reconstruction technique based on FSST4, denoted by FSST4-MR. 
An illustration of the reconstructed signal obtained with RRP-LCR-MR superimposed on the one given by the numerical relativity is displayed in 
Fig. \ref{Fig9} (b). Then, we estimate the quality of signal reconstruction with the three techniques when the input SNR varies. The results depicted in Fig. \ref{Fig9} (c) show 
that the three methods behave similarly, though RRP-MR and RRP-LCR-MR are always slightly better. 
What is very specific to the studied gravitational wave signal is that the part associated with the strong frequency modulation is very energetic, 
and a slight inaccuracy in IF estimation using the spline approximation at the locations corresponding to strong frequency modulation has a strong impact on mode reconstruction 
with RRP-LCR-MR. For this reason, it may not be that interesting to use the latter technique instead of RRP-MR in that context.
It is also worth noting here that  FSST4-MR is very specific to signals containing very oscillatory phases, which is the case of the gravitational wave when the ringdown occurs. 
For any other modes like those studied before, a lower order synchrosqueezing transform would lead to better results  \cite{meignen2021use}. 
Another limitation of techniques based on synchrosqueezing transforms is that there are not adapted to long signals 
since the reconstruction of the modes from these transforms when the hop-size is larger than one is not tractable \cite{meignen2018retrieval}. 
Finally, as FSST4-MR uses S-RD, it depends on the user defined parameter $B_f$ which is not the case of RRP-MR or RRP-LCR-MR.
These last techniques thus offer a nice alternative to the best state-of-the-art techniques for the reconstruction of very complex signals such as gravitational-wave 
signals.  
 
\section{Conclusion}
In this paper, we have introduced a novel technique to detect the ridges made by the modes of a multicomponent signal in the time-frequency plane. 
We focused on the design of a technique enabling the computation of the ridges in very noisy situations. 
For that purpose, we remarked that when the noise level is high, it is more relevant to associate with a mode ridge portions, rather than try to connect directly local maxima in the time-frequency plane as classical ridge detectors often do. This idea is the key principle to the new proposed ridge detector which 
is shown to outperform state-of-the-art methods based on time-frequency representations. As far as mode reconstruction is concerned, to combine the proposed ridge
with a local linear chirp approximation for each mode results in improved performance compare with other recent techniques, in most cases. 
Finally, the relevance of the proposed approach is also confirmed by analyzing its behavior on gravitational-wave signals. 
Some remaining limitations of the present work are that the proposed ridge detector cannot deal with crossing modes and assumes the number of modes is fixed for the whole signal duration. In a near a future, we will investigate how to adapt this  algorithm to such situations. 

\bibliographystyle{IEEEtran}
\bibliography{CRAS_new}

\begin{thebibliography}{10}
\providecommand{\url}[1]{#1}
\csname url@samestyle\endcsname
\providecommand{\newblock}{\relax}
\providecommand{\bibinfo}[2]{#2}
\providecommand{\BIBentrySTDinterwordspacing}{\spaceskip=0pt\relax}
\providecommand{\BIBentryALTinterwordstretchfactor}{4}
\providecommand{\BIBentryALTinterwordspacing}{\spaceskip=\fontdimen2\font plus
\BIBentryALTinterwordstretchfactor\fontdimen3\font minus
  \fontdimen4\font\relax}
\providecommand{\BIBforeignlanguage}[2]{{%
\expandafter\ifx\csname l@#1\endcsname\relax
\typeout{** WARNING: IEEEtran.bst: No hyphenation pattern has been}%
\typeout{** loaded for the language `#1'. Using the pattern for}%
\typeout{** the default language instead.}%
\else
\language=\csname l@#1\endcsname
\fi
#2}}
\providecommand{\BIBdecl}{\relax}
\BIBdecl

\bibitem{gribonval2003harmonic}
R.~Gribonval and E.~Bacry, ``Harmonic decomposition of audio signals with
  matching pursuit,'' \emph{IEEE Transactions on Signal Processing}, vol.~51,
  no.~1, pp. 101--111, 2003.

\bibitem{Herry2017}
C.~L. Herry, M.~Frasch, A.~J. Seely, and H.-T. Wu, ``Heart beat classification
  from single-lead {ECG} using the synchrosqueezing transform,''
  \emph{Physiological Measurement}, vol.~38, no.~2, pp. 171--187, 2017.

\bibitem{Lin2016}
Y.-Y. Lin, H.-T. Wu, C.-A. Hsu, P.-C. Huang, Y.-H. Huang, and Y.-L. Lo, ``Sleep
  apnea detection based on thoracic and abdominal movement signals of wearable
  piezoelectric bands,'' \emph{IEEE journal of biomedical and health
  informatics}, vol.~21, no.~6, pp. 1533--1545, 2017.

\bibitem{Flandrin1998}
P.~Flandrin, \emph{Time-frequency/time-scale analysis}.\hskip 1em plus 0.5em
  minus 0.4em\relax Academic Press, 1998, vol.~10.

\bibitem{Boashash2003}
B.~Boashash, \emph{Time frequency signal analysis and processing - A
  comprehensive reference}.\hskip 1em plus 0.5em minus 0.4em\relax Gulf
  Professional Publishing, 2003.

\bibitem{stankovic2014time}
L.~Stankovic, M.~Dakovic, and T.~Thayaparan, \emph{Time-frequency signal
  analysis with applications}.\hskip 1em plus 0.5em minus 0.4em\relax Artech
  house, 2014.

\bibitem{stankovic2001measure}
L.~Stankovi{\'c}, ``A measure of some time--frequency distributions
  concentration,'' \emph{Signal Processing}, vol.~81, no.~3, pp. 621--631,
  2001.

\bibitem{stankovic2001performance}
L.~Stankovic, M.~Dakovic, and V.~Ivanovic, ``Performance of spectrogram as {IF}
  estimator,'' \emph{Electronics Letters}, vol.~37, no.~12, pp. 797--799, 2001.

\bibitem{Carmona1997}
R.~Carmona, W.~Hwang, and B.~Torresani, ``Characterization of signals by the
  ridges of their wavelet transforms,'' \emph{IEEE Transactions on Signal
  Processing}, vol.~45, no.~10, pp. 2586--2590, Oct 1997.

\bibitem{meignen2018retrieval}
S.~Meignen and D.-H. Pham, ``Retrieval of the modes of multicomponent signals
  from downsampled short-time {F}ourier transform,'' \emph{IEEE Transactions on
  Signal Processing}, vol.~66, no.~23, pp. 6204--6215, 2018.

\bibitem{djurovic2004algorithm}
I.~Djurovi{\'c} and L.~Stankovi{\'c}, ``An algorithm for the {W}igner
  distribution based instantaneous frequency estimation in a high noise
  environment,'' \emph{Signal Processing}, vol.~84, no.~3, pp. 631--643, 2004.

\bibitem{Carmona1999}
R.~Carmona, W.~Hwang, and B.~Torresani, ``Multiridge detection and
  time-frequency reconstruction,'' \emph{IEEE Transactions on Signal
  Processing}, vol.~47, no.~2, pp. 480--492, Feb 1999.

\bibitem{zhu2019two}
X.~Zhu, Z.~Zhang, J.~Gao, and W.~Li, ``Two robust approaches to multicomponent
  signal reconstruction from {STFT} ridges,'' \emph{Mechanical Systems and
  Signal Processing}, vol. 115, pp. 720--735, 2019.

\bibitem{Daubechies2011}
I.~Daubechies, J.~Lu, and H.-T. Wu, ``Synchrosqueezed wavelet transforms: an
  empirical mode decomposition-like tool,'' \emph{Applied and Computational
  Harmonic Analysis}, vol.~30, no.~2, pp. 243--261, 2011.

\bibitem{laurent2020novel}
N.~Laurent and S.~Meignen, ``A novel time-frequency technique for mode
  retrieval based on linear chirp approximation,'' \emph{IEEE Signal Processing
  Letters}, vol.~27, pp. 935--339, 2020.

\bibitem{li2019adaptive}
L.~Li, H.~Cai, and Q.~Jiang, ``Adaptive synchrosqueezing transform with a
  time-varying parameter for non-stationary signal separation,'' \emph{Applied
  and Computational Harmonic Analysis}, 2019.

\bibitem{chui2016signal}
C.~K. Chui and H.~Mhaskar, ``Signal decomposition and analysis via extraction
  of frequencies,'' \emph{Applied and Computational Harmonic Analysis},
  vol.~40, no.~1, pp. 97--136, 2016.

\bibitem{li2020direct}
L.~Li, C.~K. Chui, and Q.~Jiang, ``Direct signal separation via extraction of
  local frequencies with adaptive time-varying parameters,'' \emph{arXiv
  preprint arXiv:2010.01866}, 2020.

\bibitem{chui2021analysis}
C.~K. Chui, Q.~Jiang, L.~Li, and J.~Lu, ``Analysis of an adaptive short-time
  fourier transform-based multicomponent signal separation method derived from
  linear chirp local approximation,'' \emph{Journal of Computational and
  Applied Mathematics}, p. 113607, 2021.

\bibitem{Thakur2011}
G.~Thakur and H.-T. Wu, ``Synchrosqueezing-based recovery of instantaneous
  frequency from nonuniform samples.'' \emph{SIAM J. Math. Analysis}, vol.~43,
  no.~5, pp. 2078--2095, 2011.

\bibitem{Thakur2013}
G.~Thakur, E.~Brevdo, N.~S. Fuckar, and H.-T. Wu, ``The synchrosqueezing
  algorithm for time-varying spectral analysis: robustness properties and new
  paleoclimate applications,'' \emph{Signal Processing}, vol.~93, no.~5, pp.
  1079--1094, May 2013.

\bibitem{oberlin2014}
T.~Oberlin, S.~Meignen, and V.~Perrier, ``The {F}ourier-based synchrosqueezing
  transform,'' in \emph{2014 IEEE International Conference on Acoustics, Speech
  and Signal Processing (ICASSP)}, May 2014, pp. 315--319.

\bibitem{Meignen2016a}
S.~Meignen, T.~Oberlin, P.~Depalle, P.~Flandrin, and S.~McLaughlin, ``Adaptive
  multimode signal reconstruction from time--frequency representations,''
  \emph{Phil. Trans. R. Soc. A}, vol. 374, no. 2065, p. 20150205, 2016.

\bibitem{chen2017intrinsic}
S.~Chen, Z.~Peng, Y.~Yang, X.~Dong, and W.~Zhang, ``Intrinsic chirp component
  decomposition by using {F}ourier series representation,'' \emph{Signal
  Processing}, vol. 137, pp. 319--327, 2017.

\bibitem{colominas2020fully}
M.~A. Colominas, S.~Meignen, and D.-H. Pham, ``Fully adaptive ridge detection
  based on {STFT} phase information,'' \emph{IEEE Signal Processing Letters},
  2020.

\bibitem{meignen2017demodulation}
S.~Meignen, D.-H. Pham, and S.~McLaughlin, ``On demodulation, ridge detection,
  and synchrosqueezing for multicomponent signals,'' \emph{IEEE Transactions on
  Signal Processing}, vol.~65, no.~8, pp. 2093--2103, 2017.

\bibitem{behera2018theoretical}
R.~Behera, S.~Meignen, and T.~Oberlin, ``Theoretical analysis of the
  second-order synchrosqueezing transform,'' \emph{Applied and Computational
  Harmonic Analysis}, vol.~45, no.~2, pp. 379--404, 2018.

\bibitem{Pham2018a}
D.-H. Pham and S.~Meignen, ``A novel thresholding technique for the denoising
  of multicomponent signals,'' in \emph{43th International Conference on
  Acoustics, Speech, and Signal Processing (ICASSP)}, 2018.

\bibitem{Donoho1994}
D.~Donoho and I.~Johnstone, ``Ideal spatial adaptation via wavelet shrinkage,''
  \emph{Biometrika}, vol.~81, pp. 425--455, 1994.

\bibitem{baraniuk2001measuring}
R.~G. Baraniuk, P.~Flandrin, A.~J. Janssen, and O.~J. Michel, ``Measuring
  time-frequency information content using the {R}{\'e}nyi entropies,''
  \emph{IEEE Transactions on Information theory}, vol.~47, no.~4, pp.
  1391--1409, 2001.

\bibitem{meignen2020use}
S.~Meignen, M.~Colominas, and D.-H. Pham, ``On the use of {R}{\'e}nyi entropy
  for optimal window size computation in the short-time {F}ourier transform,''
  in \emph{ICASSP 2020-2020 IEEE International Conference on Acoustics, Speech
  and Signal Processing (ICASSP)}.\hskip 1em plus 0.5em minus 0.4em\relax IEEE,
  2020, pp. 5830--5834.

\bibitem{Auger1995}
F.~Auger and P.~Flandrin, ``Improving the readability of time-frequency and
  time-scale representations by the reassignment method,'' \emph{IEEE
  Transactions on Signal Processing}, vol.~43, no.~5, pp. 1068--1089, 1995.

\bibitem{li2020adaptive}
L.~Li, H.~Cai, H.~Han, Q.~Jiang, and H.~Ji, ``Adaptive short-time {F}ourier
  transform and synchrosqueezing transform for non-stationary signal
  separation,'' \emph{Signal Processing}, vol. 166, p. 107231, 2020.

\bibitem{pham2017high}
D.~H. Pham and S.~Meignen, ``High-order synchrosqueezing transform for
  multicomponent signals analysis-with an application to gravitational-wave
  signal.'' \emph{IEEE Trans. Signal Processing}, vol.~65, no.~12, pp.
  3168--3178, 2017.

\bibitem{Abbott2016}
B.~P. Abbott, R.~Abbott, T.~Abbott, M.~Abernathy, F.~Acernese, K.~Ackley,
  C.~Adams, T.~Adams, P.~Addesso, R.~Adhikari \emph{et~al.}, ``Observation of
  gravitational waves from a binary black hole merger,'' \emph{Physical review
  letters}, vol. 116, no.~6, p. 061102, 2016.

\bibitem{abbott2016gw151226}
B.~P. Abbott, R.~Abbott \emph{et~al.}, ``{GW151226}: Observation of
  gravitational waves from a 22-solar-mass binary black hole coalescence,''
  \emph{Physical Review Letters}, vol. 116, no.~24, p. 241103, 2016.

\bibitem{meignen2021use}
S.~Meignen, D.-H. Pham, and M.~A. Colominas, ``On the use of short-time fourier
  transform and synchrosqueezing-based demodulation for the retrieval of the
  modes of multicomponent signals,'' \emph{Signal Processing}, vol. 178, p.
  107760, 2021.

\end{thebibliography}

\ifCLASSOPTIONcaptionsoff
  \newpage
\fi

\end{document}